%% file: single_document.tex
\documentclass[11pt,a4paper]{article}
\usepackage{jheppub}
\usepackage{atlasphysics}
\usepackage{subfigure}
\usepackage{mathrsfs}
\usepackage{url} 
\usepackage{xspace}
\usepackage{dcolumn}
\usepackage{booktabs}
\usepackage{afterpage}
\usepackage{rotating}
\usepackage{multirow}
\usepackage{array}
\usepackage{float}
\usepackage{placeins}
\usepackage{ae}

\newcolumntype{C}[1]{>{\centering\arraybackslash}p{#1}}

\usepackage[T1]{fontenc} 

\newcommand{\alpgen}{{\sc alpgen}}

\newcommand{\jimmy}{{\sc jimmy}}
\newcommand{\mcnlo}{{\sc mc@nlo}}
\newcommand{\powheg}{{\sc powheg}}
\newcommand{\powhegwith}[1]{\powheg {\sc (#1)}}

\newcommand{\acermc}{{\sc acermc}}
\newcommand{\pythia}{{\sc pythia}}

\newcommand{\herwig}{{\sc herwig}}

\newcommand{\geant}{{\sc geant4}}
\newcommand{\tauola}{{\sc tauola}}
\newcommand{\photos}{{\sc photos}}
\newcommand{\mcfm}{{\sc mcfm}}

\newcommand{\cteq}{{\sc cteq6l1}}
\newcommand{\cteqfive}{{\sc cteq5l}}
\newcommand{\cteqsixsix}{{\sc cteq66}}
\newcommand{\nnpdf}{{\sc nnpdf~2.0}}
\newcommand{\ctten}{{\sc ct10nlo}}
\newcommand{\hera}{{\sc herapdf15nlo}}
\newcommand{\mstw}{{\sc mstw2008nlo}}
\newcommand{\toppp}{{\sc top++}}

\newcommand{\wjets}{\ensuremath{W\!+\!\mathrm{jets}}\xspace}
\newcommand{\wpjets}{\ensuremath{{W^+}\!+\!\mathrm{jets}}\xspace}
\newcommand{\wnjets}{\ensuremath{{W^-}\!+\!\mathrm{jets}}\xspace}

\newcommand{\zjets}{\ensuremath{Z\!+\!\mathrm{jets}}\xspace}

\newcommand{\mtw}{\ensuremath{m_\mathrm{T}(W)}\xspace}

\def\ejets{\ensuremath{e+\textrm{jets}}\xspace}

\def\mujets{\ensuremath{\mu+\textrm{jets}}\xspace}

\def\ppp2011c{\powheg+\pythia Perugia2011 \ttbar\xspace}
\def\pt{\ensuremath{p_{\mathrm{T}}}\xspace} 
\def\pT{\ensuremath{p_{\mathrm{T}}}\xspace} 
\def\pTx{\ensuremath{p_{\mathrm{T}}}}

\def\Nbgnd{\ensuremath{N_\textrm{bgnd}}}
\def\frnp{\ensuremath{f_\textrm{reco!part}}}
\def\fmisassign{\ensuremath{f_\mathrm{misassign}}}
\def\fpnr{\ensuremath{f_\textrm{part!reco}}}
\def\Mres{\ensuremath{\mathbf{M}_\textrm{part}^\textrm{reco}}}

\def\Nparti{\ensuremath{N_\textrm{part}^{i}}}
\def\fpartrecoi{\ensuremath{f_\textrm{part!reco}}^{i}}
\def\frecopartj{\ensuremath{f_\textrm{reco!part}}^{j}}
\def\mrecopartij{\ensuremath{\mathbf{M}_{\textrm{reco,}{j}}^{\textrm{part,}{i}}}}

\def\xrecoj{\ensuremath{x_\textrm{reco}^{j}}}
\def\xparti{\ensuremath{x_\textrm{part}^{i}}}

\def\Nbgndj{\ensuremath{N_\textrm{bgnd}^{j}}}
\def\Nparti{\ensuremath{N_\textrm{part}^{i}}}
\def\Nrecoj{\ensuremath{N_\textrm{reco}^{j}}}
\def\fmisassignj{\ensuremath{f^{j}_\mathrm{misassign}}}

\newcommand{\pseudotop}{\mbox{$\hat{t}$}}
\newcommand{\pseudotopl}{\mbox{$\pseudotop_{\rm{l}}$}}
\newcommand{\pseudotoph}{\mbox{$\pseudotop_{\rm{h}}$}}

\begin{document}
\author{The ATLAS Collaboration}
\title{Differential top--antitop cross-section measurements as a function of observables constructed from final-state particles using pp collisions at $\sqrt{s}=7$ TeV in the ATLAS detector}


\abstract{
Various differential cross-sections are measured in top-quark pair (\ttbar) events produced in 
proton--proton collisions at a centre-of-mass energy of $\sqrt{s} = 7$ TeV at the LHC with the ATLAS detector. 
These differential cross-sections are presented
in a data set corresponding to an integrated luminosity of $4.6$~fb$^{-1}$. 
The differential cross-sections are presented in terms of kinematic variables, 
such as momentum, rapidity and invariant mass, of a top-quark proxy 
referred to as the pseudo-top-quark as well as the pseudo-top-quark pair system.
The dependence of the measurement on theoretical models is minimal.

The measurements are performed on \ttbar{} events in the lepton+jets channel, requiring exactly one 
charged lepton and at least four jets with at least two of them tagged as originating from a $b$-quark. 
The hadronic and leptonic pseudo-top-quarks are defined via the leptonic or hadronic
decay mode of the $W$ boson produced by the top-quark decay in events with a single charged lepton.
%
Differential cross-section measurements of the pseudo-top-quark variables are compared
with several Monte Carlo models that implement next-to-leading order
or leading-order multi-leg matrix-element calculations.
}

\maketitle
\newpage

\section{Introduction}
\label{sec:introduction}

The large number of top-quark pair ($\ttbar$) events produced at the Large Hadron Collider (LHC) 
has allowed the ATLAS~\cite{Aad:2008zzm} and CMS~\cite{Chatrchyan:2008aa} experiments to perform  
precise inclusive and differential top-quark related cross-section measurements. 
Both experiments have recently published measurements of the inclusive \ttbar\ production cross-section in
proton--proton ($pp$) collisions at centre-of-mass energies, $\sqrt{s}$, of 7 and 
8~\TeV\ ~\cite{Aad:2014kva,Aad:2014jra,ATLAS:2012aa,Aad:2012mza,Aad:2012vip,Chatrchyan:2013ual,Chatrchyan:2013kff,Chatrchyan:2013faa,
Chatrchyan:2012ria,Chatrchyan:2012bra,Chatrchyan:2012vs,Chatrchyan:2011yy} as well as differential
cross-section measurements as functions of the top-quark transverse momentum (\pT) and rapidity ($y$), and of the mass ($m_{t\bar{t}}$) 
and $y$ of the \ttbar\ system~\cite{Aad:2014zka,Aad:2012hg,Chatrchyan:2012saa}. 
These cross-section measurements triggered recent work on Quantum Chromodynamics (QCD) calculations of heavy quark 
production \cite{Guzzi:2014wia,Catani:2014qha,Campbell:2014kua,Kidonakis:2014pja,Ferroglia:2013awa,Dowling:2013baa}.

Precision measurements of \ttbar\ production provide the opportunity to conduct tests of 
predictions based on perturbative QCD 
and gain direct information on the gluon parton distribution function (PDF) at large momentum fractions ($x_{\rm Bj}$) of about 
0.1-0.5~\cite{Czakon:2013tha}. 
Differential \ttbar{} measurements are particularly sensitive to 
the \ttbar{} production mechanism in QCD in a region of 
parton momentum fractions at large momentum transfers.
Such studies can lead to improvements in background predictions for Higgs measurements and searches for physics beyond the Standard Model (SM).

In the SM, a top-quark decays to a $W$ boson and a $b$-quark with a
branching fraction close to unity. 
Hence there are three \ttbar\ signatures
that correspond to different decay modes of the $W$ bosons. 
The signal for this study is in the single-lepton channel. 
It corresponds to the case where one $W$ boson
decays directly, or via an intermediate $\tau$ decay, into an electron or muon and at least one neutrino, and the other into a pair of quarks.
The neutrino(s) will escape the detector unseen, leading to missing transverse
momentum whose magnitude is denoted by \met.

The model dependence \ttbar{} differential cross section measurements
presented at the level of top quarks, i.e. corrected for parton shower effects
and hadronisation, has been an ongoing concern.
This paper presents differential \ttbar\ cross-section measurements using a  
definition where the variables are constructed from 
an object that is directly related to detector-level observables. This top-quark proxy 
object is referred to as the $pseudo$-$top$-$quark$ (\pseudotop). 
The goal of presenting measurements using a definition where the variables are
constructed from reconstructed charged lepton, jet and missing
transverse momentum objects, is to allow precision tests of QCD in final states with top-quarks,
using reconstructed objects that avoid large model-dependent extrapolation corrections to the
parton-level top-quark but remain 
well correlated with corresponding objects reconstructed from the partons. 

In this approach, corrections applied to data 
depend less on Monte Carlo (MC) models that describe the hard scattering process 
modelled by matrix elements calculated to a given order, 
the emission of additional partons from the hard scattering process, 
parton shower effects, hadronisation and 
multiple-parton
interactions.
In particular, at low top-quark \pt{}, the modelling of soft parton 
emissions can significantly modify the kinematics of the top-quark pair.
Observables based on stable particles can be unambiguously compared
to MC generator predictions and therefore provide useful benchmarks 
to test and further develop MC models and to adjust free model parameters.

The \pseudotop{} object can be evaluated for  
hadronic or leptonic decays of the top-quark from the detector information
or analogously from the stable final-state particles generated by MC simulations.
The differential cross-sections are measured
as functions of the transverse momentum \pTx(\pseudotop) and rapidity $y(\pseudotop)$ 
of the leptonic (\pseudotopl) and the hadronic (\pseudotoph) pseudo-top-quark
as well as the 
transverse momentum $\pT(\pseudotopl \pseudotoph)$, rapidity $y(\pseudotopl \pseudotoph)$ 
and invariant mass $m(\pseudotopl \pseudotoph)$ of the 
reconstructed \ttbar{} system (\pseudotopl \pseudotoph).

This paper is structured as follows. 
The definition of \pseudotop{} together with the detector-level or particle-level
objects used to construct \pseudotop{} are presented in section \ref{sec:meas-def}. 
The correlation in MC simulation  
between reconstructed \pseudotop{} observables and corresponding top-quark observables at the parton level is also discussed in this section.
Section~\ref{sec:atlas} provides a short overview of the ATLAS detector. 
A description of the different MC samples in the study is found in section \ref{sec:monte-carlo-samples}. 
The data and MC event selection is described in section~\ref{sec:data_event_sel} together with the reconstruction of final-state objects. Section~\ref{sec:systematic_uncertainties} covers
the treatment and evaluation of systematic uncertainties. Comparisons between data and MC simulation, for the yields
and pseudo-top-quark distributions before unfolding, are presented in section~\ref{sec:reco}. A description of the unfolding,
the results obtained for data and various MC models and concluding remarks are found in sections~\ref{sec:corr_chancomb}, \ref{sec:results}
and \ref{sec:conclusions}, respectively.

\section{Measurement definition}
\label{sec:meas-def}

The model dependence of parton-level \ttbar{} cross section measurements
has been an ongoing concern for QCD studies in particular in events with top quarks. 
The use of particle-based definitions in cross-section measurements is a standard
methodology in high-energy physics \cite{Buttar:2008jx,Moch:2014tta} to reduce the model dependence. 
Standardised tools exist to compare theory predictions for such measurements \cite{Bromley:1995np,Waugh:2006ip,Buckley:2010ar}. 
In \ttbar{} events, such measurements were published for the inclusive cross-section \cite{Aad:2014kva,Aad:2014jra} 
and for differential cross-section measurements as a function of the transverse momentum 
and the rapidity of the final-state leptons and jets \cite{ATLAS:2012al,Aad:2014iaa,Chatrchyan:2012saa,Chatrchyan:2014gma}.
In this paper the concept of the particle-based cross-section definition 
is extended to the kinematic properties of the top-quark decay products
that when correctly combined are closely related to the kinematic properties of the top-quark.
An operational definition \pseudotop{} 
that is defined from measured observables
as described below has been introduced to reduce the model dependence
of the measurements.\footnote{
Discussions related to an operational definition have occurred between 
experimentalists and theorists in the Top Physics LHC working group.}
This operational definition should result in a good correlation between the \pseudotop{} object 
and the top parton for a given Monte Carlo generator. 
However, it is not necessary that the algorithms lead to the best possible correlation
to the parton-level kinematics or the best signal to background ratio.

The differential cross-section measurements in this paper are presented in terms of the kinematics of 
the \pseudotop{} object, introduced in section~\ref{sec:introduction}.
The identification of reconstructed charged lepton, jet and \met\ objects from actual or simulated detector signals
is discussed in section~\ref{object-selection}. The identification of particle-level objects for MC events is discussed in section~\ref{sec:particle-objects}. The kinematic
fiducial region for both the reconstructed and particle-level objects is defined in section~\ref{sec:kinematic-range}.   
The algorithm used to construct leptonic or hadronic \pseudotop{} objects from either reconstructed detector-level objects or from particle-level objects is described in section~\ref{sec:hl-pseudo-top}.

The measurements presented in this paper can be directly compared to MC simulations 
using matrix-element calculations for the hard scattering, 
interfaced with parton shower and hadronisation models. 
These are referred to as particle-level predictions.
For comparisons to fixed-order QCD calculations, corrections 
for the transition from partons to hadrons
need to be applied.

\subsection{Particle objects}
\label{sec:particle-objects}

In the case of MC simulation, objects can be identified at the particle level. Leptons and jets are defined using particles 
with a mean lifetime $\tau  >  3 \times 10^{-11}$~s that are directly produced in $pp$ interactions or from subsequent
decays of particles with a shorter lifetime. 
The lepton definition only includes
prompt electrons, muons and neutrinos not originating from hadron decays
as well as electrons, muons and neutrinos from tau decays.
The electron and muon four-momenta  
are calculated after the addition of any photon four-momenta,  
not originating from hadron decay.
The electron and muon four-momenta are defined to include any photons
not originating from hadron decays that are found within $\Delta R  = \sqrt{(\Delta\phi)^2+(\Delta \eta )^2} < 0.1$
with respect to the lepton direction.
The direction of the photons is required to be within $\Delta R  = \sqrt{(\Delta\phi)^2+(\Delta \eta )^2} = 0.1$ with respect to the lepton direction
\footnote{ATLAS uses a right-handed coordinate system with its origin at the nominal interaction point (IP) in the centre of the detector and the $z$-axis along the beam pipe. The $x$-axis points from the IP to the centre of the LHC ring, and the $y$-axis points upward. Cylindrical coordinates $(r,\phi)$ are used in the transverse plane, $\phi$ being the azimuthal angle around the beam pipe. The pseudorapidity is defined in terms of the polar angle $\theta$ as $\eta=-\ln\tan(\theta/2)$.}.
The missing transverse momentum vector 
and its associated azimuthal angle are evaluated from the sum of the neutrino 
four-momenta, where all neutrinos from $W$ boson and $\tau$ decays are included.  Jets are defined by the anti-$k_t$ algorithm~\cite{Cacciari:2008gp} with a radius parameter of 0.4.  The jets include all stable particles except for the selected electrons, muons and neutrinos, 
and the photons associated with these electrons or muons.  
The presence of one or more $b$-hadrons with $\pT > 5$~\GeV\ 
associated to a jet defines it as a $b$-jet.  
To perform the matching  
between $b$-hadrons and jets, the $b$-hadron energy is scaled to a negligible value 
and included in the jet clustering
($ghost$-$matching$)~\cite{Cacciari:2008gn}.

\subsection{Kinematic range of objects}
\label{sec:kinematic-range}

The cross-section measurement is defined 
in a kinematic region where the reconstructed physics objects have a high reconstruction efficiency
(fiducial region).  The kinematic region is chosen such 
that the kinematic selections 
of the physics objects reconstructed in the detector and of the particle objects
are as close as possible. 
The fiducial region is defined in the same way
for reconstructed physics and particle objects. However, on detector-level some
additional selections can be applied.

Electrons, muons and jets are required to satisfy $\pT > 25$~\GeV\ and $|\eta| < 2.5$. The fiducial volume is defined by requiring exactly
one muon or electron, four or more jets of which at least two are $b$-jets, $\met > 30$~\GeV\ and a $W$ boson transverse mass $\mtw > 35$~\GeV.\footnote{The $W$ boson transverse mass \mtw\ is defined as $\sqrt{2 \pT^{\ell} \pT^{\nu} (1 - \cos(\phi^{\ell} - \phi^{\nu}))}$, where $\ell$ and $\nu$ refer to the charged lepton ($e$ or $\mu$) 
and the missing transverse momentum vector, respectively.
The symbol $\phi$ denotes the azimuthal angle of the lepton or missing transverse momentum vector.}
Events are discarded if the electron or muon is within $\Delta R = 0.4$ of a jet, or two jets are within $\Delta R = 0.5$ of each other.  

\subsection{Hadronic and leptonic pseudo-top-quark definition}
\label{sec:hl-pseudo-top}

The definition of \pseudotop{} as a hadronic or leptonic object is determined by the decay of  the $W$ boson. 
In the cross-section definition used in this paper,
the two highest \pT\ $b$-jets are assumed to be the $b$-jets from the top-quark decay. 

\begin{itemize}
\item In the case of \pseudotopl, the leptonically decaying 
$W$ boson is constructed from the electron or muon and the \met. The $b$-jet 
with the smallest angular separation ($\Delta R$)
from the electron or muon is then assigned as a decay product of \pseudotopl. Using 
the measured $W$ boson mass, $m_{W} = 80.399$~GeV~\cite{Nakamura:2010zzi}, and the components of 
the missing transverse momentum vector (denoted as $p_{{x},\nu}, p_{{y},\nu}$) 
associated with the $W$ boson decay neutrino, 
the $p_{{z},\nu}$ of the neutrino can be constrained:
\[ (E_{\ell}+E_{\nu})^2 - (p_{{x},\ell} + p_{{x},\nu})^2 - (p_{{y},\ell} + p_{{y},\nu})^2 - (p_{{z},\ell} + p_{{z},\nu})^2 = m_{W}^2, \]

where the subscript $\ell$ refers to the electron or muon.
Neglecting the neutrino mass, the $p_{{z},\nu}$  
of the neutrino is taken from the solution of the resulting quadratic equation:
\[ p_{{z},\nu} = \frac{-b \pm \sqrt{b^2 - 4ac}}{2a}, \]
where
\[ a = E_{\ell}^2 - p_{z,\ell}^2 \qquad b = -2 k p_{z,\ell} \qquad c = E_{\ell}^2 p_{\mathrm{T},\nu}^2 - k^2 \]
and
\[k = \frac{m_{W}^2 - m_{\ell}^2}{2} + (p_{{x},\ell}p_{{x},\nu} + p_{{y},\ell} p_{{y},\nu}). \]
If both solutions are real, the solution with the smaller
|$p_{{z},\nu}$| 
is chosen. In cases 
where $(b^2 - 4ac)$ is less than zero,  
$p_{{z},\nu}$
is taken as:
\[ p_{{z},\nu} = -\frac{b}{2a}. \]
Given the value of  $p_{{z},\nu}$, \pseudotopl{} 
is formed from the combination of 
 the charged lepton, neutrino and assigned $b$-jet. 
\item In the case of \pseudotoph, the hadronically decaying $W$ boson is constructed from the remaining two highest-\pT\ jets. The \pseudotoph{} is then defined from the hadronically decaying $W$ boson candidate and the remaining $b$-jet. 
\end{itemize}

Once the leptonic and hadronic \pseudotop{} are defined, their four-momenta can be evaluated, 
and used in the measurement to define, 
for example the transverse momentum ($\pT(\pseudotopl)$ or  $\pT(\pseudotoph$))
the rapidity ($y(\pseudotopl)$ or $y(\pseudotoph)$) and
the  transverse momentum ($\pT(\pseudotopl \pseudotoph)$), rapidity ($y(\pseudotopl \pseudotoph)$) 
and invariant mass ($m(\pseudotopl \pseudotoph)$) of the 
reconstructed \ttbar{} system.

In summary, three different definitions are used in the following sections:

\begin{itemize}
\item A \textit{parton-level top-quark} is the MC generator-level top-quark selected before it decays but after any radiative emissions;\footnote{
The four-momenta of the top-quarks produced in the hard scattering $2 \to 2$ process are modified
after gluon emission 
such that an event record with consistent information and conserved energy and momentum balance is obtained.
Technically, 
  the parton level is defined as status code 155 for \herwig{} and status code 2 for \pythia.}
\item A \textit{particle-level pseudo-top-quark (hadronic and leptonic)} is defined by stable generator-level particles within the described acceptance;
\item A \textit{detector-level pseudo-top-quark (hadronic and leptonic)} is evaluated with the use of physics objects reconstructed from detector measurements as discussed in section~\ref{object-selection}.
\end{itemize}

Detector-level distributions of \pseudotop{} variables are corrected for detector efficiency
and resolution effects to allow comparison with distributions of equivalent particle-level \pseudotop{} variables (see section~\ref{sec:corr_chancomb}).
Therefore any MC model that simulates the final-state particles from $pp$ collisions
can be compared to these data. This makes existing or future MC model comparisons possible,
independent of the presence of top-quark partons in the MC event record.
For comparisons of data to simulation it is not necessary to use the \ttbar{}
kinematic reconstruction method with the best performance, but it is important
that the method is well defined and the same definition is applied to data and
simulation.

\subsection{Kinematic comparison of the parton-level top-quark with the pseudo-top-quark}
\label{sec:kin-pseudo-top}

\begin{figure}  
\centering
\includegraphics[width=.60\textwidth]{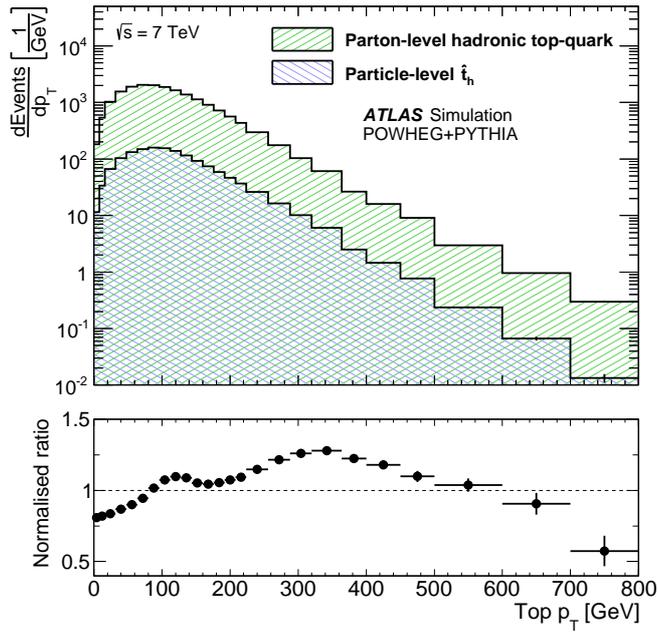}
 \caption{ 
Simulated event distribution of
the parton-level top-quark \pt{} distribution for all events  (green),  
and the particle-level pseudo-top-quark \pt{} (\pseudotoph)
for events within the fiducial region (blue). 
In both cases the top-quark that decays hadronically is chosen.
The distributions are evaluated for the same event sample
based on \powheg+\pythia{} at $\sqrt{s} = 7$~TeV. 
The upper figure is made for an arbitrary integrated luminosity.
The lower figure shows the ratio of the particle-level \pseudotoph{}
over the parton-level top-quark normalised distributions
to emphasise the difference in shape between the two.} 
\label{fig:truth-top-pt}
\end{figure}
The pseudo-top-quark definition is chosen such that it is closely related to the
top-quark parton provided by QCD calculations.

Figure \ref{fig:truth-top-pt} compares the parton-level \pt{} distribution of 
the top-quark with the particle-level \pt{} distribution of the pseudo-top-quark. 
The two distributions are shown  for the same integrated luminosity, 
using the \powheg{}+\pythia{} Monte Carlo generator (see section \ref{sec:monte-carlo-samples}). 
The number of events with a reconstructed \pseudotop{} is much smaller 
than the total number of generated \ttbar{} events. 
This is primarily due to the requirement of a single lepton, four jets
and the pseudo-top construction efficiency.

\begin{figure}[h] 
\centering
\subfigure[\label{fig:truth_parton_vs_pseudo}]{\includegraphics[width=.495\textwidth]{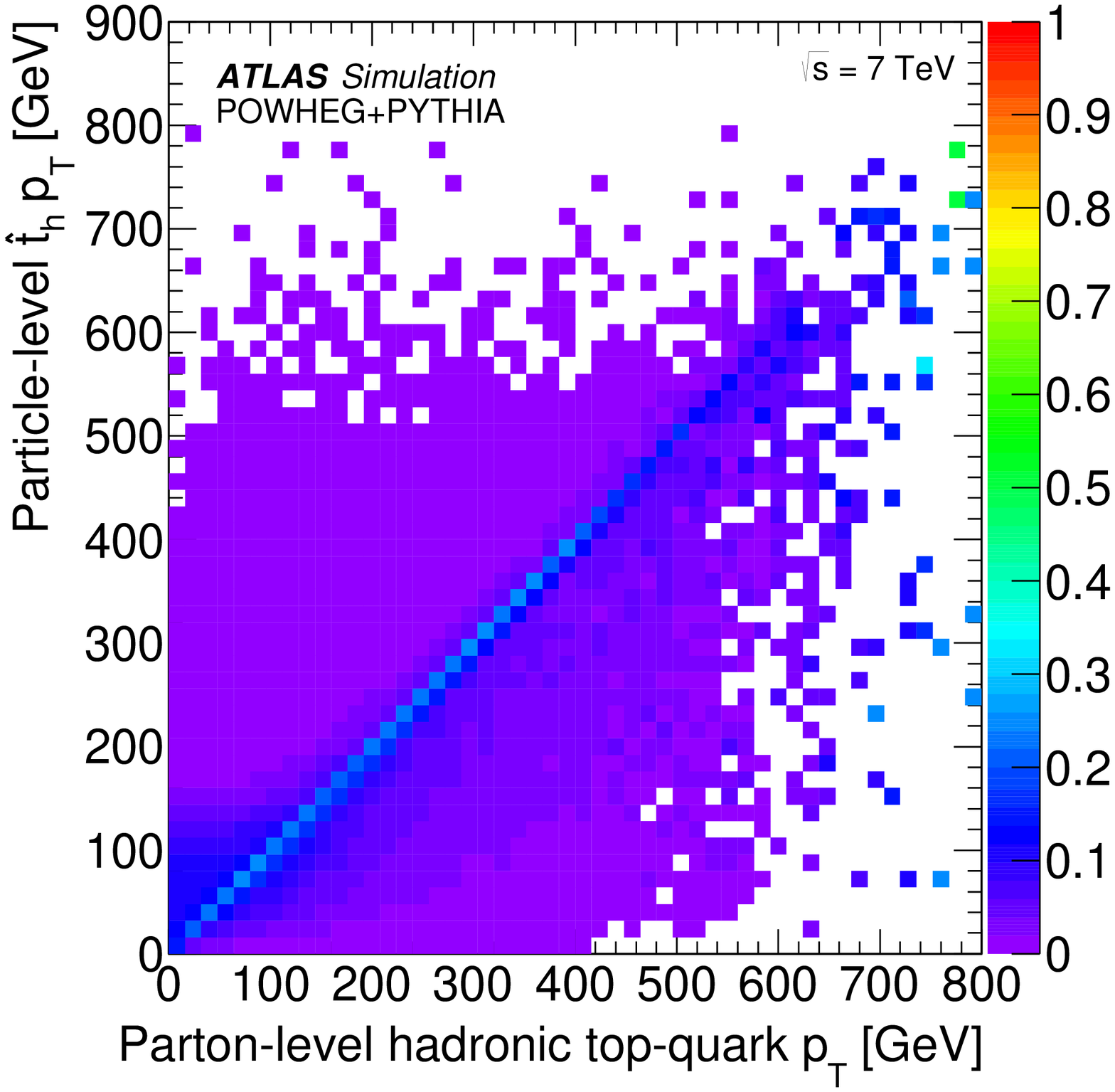}}
\subfigure[\label{fig:truth_pseudo_vs_reco}]{\includegraphics[width=.495\textwidth]{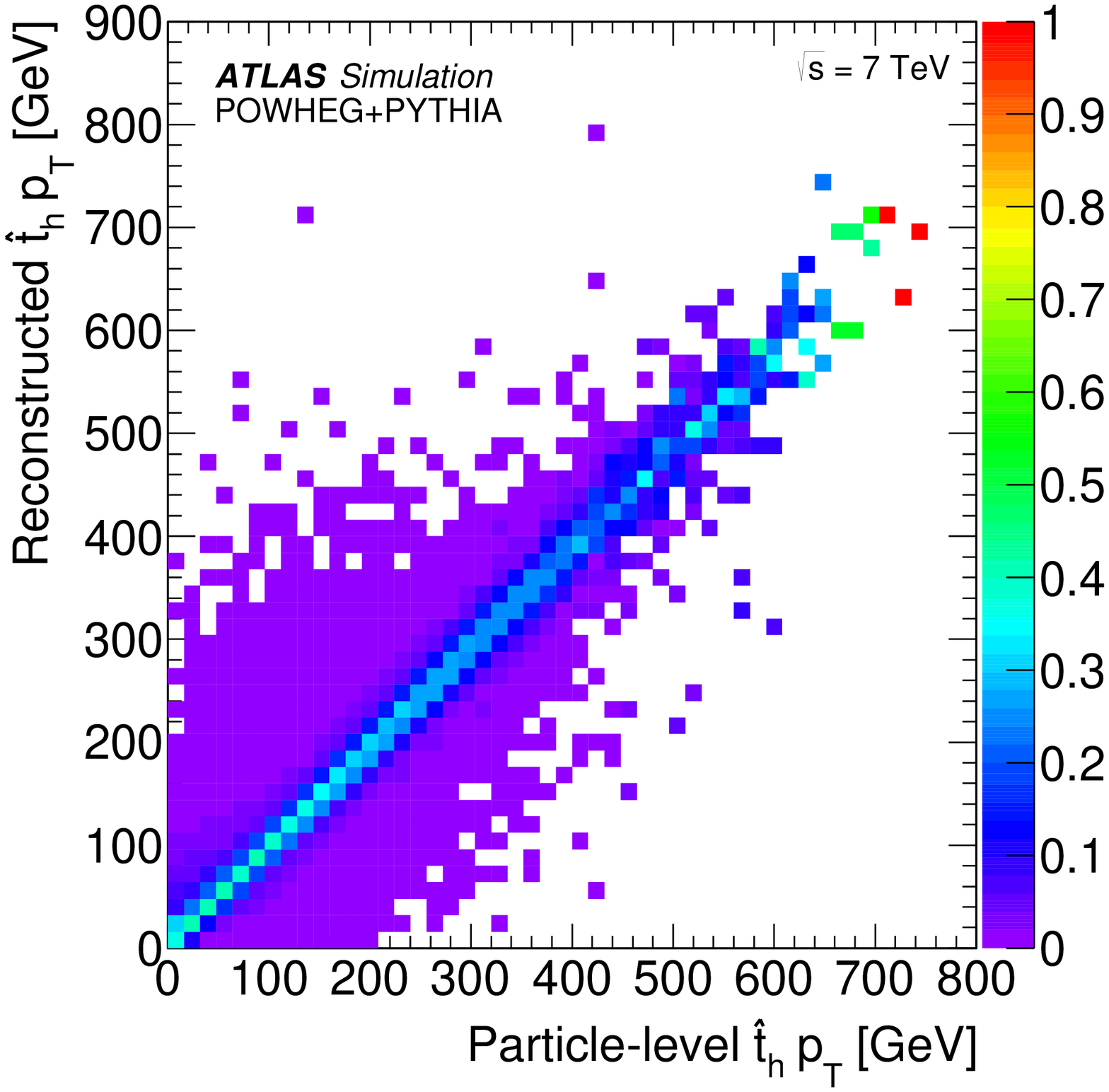}}
  \caption{ 
(a) Monte Carlo study using the nominal \powheg+\pythia{} MC sample showing 
the correlation between the parton-level top-quark \pt{} and the particle-level
 hadronic pseudo-top-quark \pt{} and (b) the correlation 
 between the particle-level hadronic pseudo-top-quark \pt{} and the hadronic 
pseudo-top-quark \pt{} evaluated from reconstructed objects. 
In each case the correlation is normalised to all the events within a bin on the horizontal axis.
}
\label{fig:truth-toppt-corr}
\end{figure}

The ratio of the normalised distributions illustrates the shape difference. 
The larger phase space of the parton-level distribution results in a softer \pt{} distribution when 
compared with the pseudo-top-quark distribution where fiducial cuts are applied. 
In addition, with increasing \pt{}, 
the decay products tend to be more collimated due to their large boost 
resulting in a reduced pseudo-top reconstruction efficiency.
Using the same Monte Carlo sample, figure \ref{fig:truth-toppt-corr}(a) shows the correlation between the generated parton-level \pt{} distribution and the particle-level hadronic
pseudo-top-quark \pt{}. 
About $50\%$ of the events show a strong correlation between the top quark parton and the particle-level \pseudotoph.
 These are the cases where the correct combination of the final state object was found to reconstruct the particle-level \pseudotoph. 
Despite the 
good correlation, there remains a significant bin migration in the region populated by the majority of top-quarks. 
These migrations can be affected by a change in the modelling of \ttbar{} production. This makes any measurement extrapolated to the parton level model-dependent. 
However, there is a much stronger correlation between the
the hadronic particle-level and detector-level pseudo-top-quark \pt{} distributions
as shown in figure \ref{fig:truth-toppt-corr}(b). Hadronic pseudo-top-quark measurements that are presented at the particle level 
are therefore less affected by model-dependent corrections. 
 
\section{ATLAS detector}
\label{sec:atlas}

The ATLAS detector~\cite{Aad:2008zzm} is a general-purpose detector
 that covers nearly the entire solid angle around one of the $pp$ interaction points of the LHC~\cite{Evans:2008zzb}.  It is composed of an inner tracking 
detector (ID), covering a range of $|\eta| < 2.5$, surrounded by a superconducting solenoid that provides a 2~T magnetic field, high-granularity  electromagnetic (EM) and hadronic sampling calorimeters and a muon spectrometer (MS) that incorporates a system of three air-core superconducting toroid magnets, each with eight coils.  The ID comprises a silicon pixel detector, a silicon microstrip detector (SCT), and a transition radiation tracker (TRT).  The EM barrel calorimeter is composed of a liquid-argon (LAr) active medium and lead absorbers.  The hadronic calorimeter is constructed from steel absorber and scintillator  
tiles in the central pseudorapidity range of $|\eta| < 1.7$, whereas the end-cap and forward regions are instrumented with LAr 
calorimeters for both the electromagnetic and hadronic energy measurements up to $|\eta| = 4.9$. The MS toroid magnets are arranged with an eight-fold azimuthal coil symmetry around the calorimeters. 
Three layers of muon spectrometer chambers surround the toroids. High-precision drift tubes and, at small radius in the end-cap, region cathode strip chambers provide an independent 
momentum measurement.  Resistive plate chambers in the central region and fast thin gap chambers in the end-cap region provide a muon trigger.

Data are selected from inclusive $pp$ interactions using a three-level trigger system.  A hardware-based first-level trigger is used to initially reduce the trigger rate to approximately 75~kHz.  The detector readout is then available for two stages of software-based (higher-level) triggers.  In the second level, partial object reconstruction is carried out to improve the selection and at the last level, the event filter, a full online event reconstruction is made to finalise the event selection. During the 2011 run period, the selected event rate for all triggers following the event filter was  approximately 300~Hz.

\section{Monte Carlo simulation}
\label{sec:monte-carlo-samples}

Monte Carlo simulation samples were generated to correct the measurements for detector effects.
The production of \ttbar\ events is modelled using the \powheg{}~\cite{Alioli:2010xd,Frixione:2007vw,Frixione:2007nw,Nason:2004rx},  
\mcnlo{}~\cite{Frixione:2002ik} and \alpgen{}~\cite{MAN-0301} generators. \powheg{} and \mcnlo{} use NLO matrix-element calculations interfaced with parton showers
and \alpgen{} implements a leading-order (LO) multi-leg calculation for up to five additional partons, 
which are subsequently matched to parton showers~\cite{Mangano:2001xp}. 
A common feature of the samples, unless stated otherwise, is the generation of the underlying event 
and parton shower, which is either performed 
by \pythia{} (v6.425)~\cite{Sjostrand:2006za} or by \herwig{} (v6.520)~\cite{COR-0001} together with \jimmy{} (v4.31)~\cite{JButterworth:1996zw}. The program
 \tauola{}~\cite{tauola} is used for the decays of $\tau$-leptons and \photos{}~\cite{photos} for photon radiation. 

A \powheg{} sample was generated using the version \powheg{}-hvq p4~\cite{Frixione:2007nw}, with 
the \ctten{} PDF set~\cite{CT10Lai2010} and the default factorisation and renormalisation scales 
set to $Q^2= m_t^2 + \pT^2$, where $m_t$ is the top-quark mass and \pT{} the top-quark transverse momentum as evaluated
for the underlying Born configuration (i.e. before radiation).
The \powheg{} matrix-element calculation was interfaced
with \pythia{}, using the ``C''  variant of the Perugia 2011 tunes~\cite{Skands:2010ak} and the corresponding \cteq{} PDF set~\cite{Stump:2003yu}. This sample
is referred to as ``\powheg{}+\pythia{}'' and is the benchmark signal sample in this study. 

An additional \powheg{} sample was generated 
in order to compare the parton shower and
fragmentation models. It was interfaced with \herwig{} together 
with \jimmy{} for the underlying event model using the AUET2 tune~\cite{PUB-2010-014} (``\powheg{}+\herwig{}''). 
To evaluate the importance of the gluon PDF on the final corrected distributions, an alternative \powheg{} sample was produced 
 with the \hera{} PDF set~\cite{HERAPDF15} 
(``\powhegwith{herapdf}+\pythia{}'').  
This PDF set is based on HERA~I data together with the inclusion of the precise high-$Q^2$ preliminary HERA~II data. Simulations using this PDF set are in 
good agreement with early $W$ and $Z$ boson production measurements at the LHC~\cite{Aad:2011dm}. 

To compare the \powheg{} predictions, alternative samples were generated using \mcnlo{} (v4.01) with the \ctten{} PDF set and its default renormalisation and factorisation scales: 
$Q^2=((p_{{\rm T},t}^2 + p_{{\rm T},\bar{t}}^2)/2) + m_{t}^2$.
\mcnlo{} was interfaced to \herwig{} and \jimmy{}, using the AUET2 tune 
(``\mcnlo{}+\herwig{}''). 

To assess the modelling of LO matrix-element calculations for additional partons, 
the \alpgen{}~\cite{MAN-0301} generator (v2.13) was 
used together with the \cteq{} PDF set and associated strong coupling constant, $\alphas\left(m_{Z}\right) = 0.129$. 
The produced processes correspond to the LO matrix elements for 
\ttbar\ with
five inclusive associated partons and $t\bar{t}+b\bar{b}$ 
and $t\bar{t}+c\bar{c}$ states, at its
default renormalisation scale and with the factorisation scale
set to $Q^2=\sum m^2+\pt^2$.  Here the sum runs over heavy quarks and light 
jets\footnote{Defined to be a jet comprising gluons and light quarks and subsequently used in the MLM matching scheme.} with mass $m$ and transverse momentum $\pt$.
The \alpgen{} samples were interfaced with \herwig{} and \jimmy{}, using the 
MLM parton--jet matching scheme~\cite{MAN-0301} with a matching scale of 20~\GeV. The exclusive heavy-flavour samples were combined with the inclusive samples, after the removal of overlapping events. This sample is referred to as ``\alpgen{}+\herwig{}''. 

Both the NLO matrix-element-based MC models and the LO multi-leg MC models 
have higher-order correction uncertainties that can be 
estimated in terms of initial-state radiation (ISR) and final-state 
radiation (FSR) variations. \alpgen{} (v2.14) is used to generate \ttbar\ samples with the \cteqfive{} PDF set~\cite{CTEQ5}, the \pythia{} parton shower and 
the Perugia 2011 tune.
Nominal and shifted ISR/FSR samples were produced 
with an $\alphas$ corresponding to $\Lambda_{\rm QCD}=0.26$ GeV as used in the
Perugia tune and by modifying the renormalisation scale at each local vertex 
in the matrix element 
by a factor of 2.0 (0.5) relative to the original scale 
to obtain more (less) radiation.
The renormalisation scale is varied
by the same factor in the matrix-element calculation
and \pythia{}, where the radHi and radLo Perugia 2011 \pythia{} tunes are used~\cite{Skands:2010ak}.  
These samples are referred to as \alpgen{}+\pythia{}(\alphas\ {\small Up}) 
and \alpgen{}+\pythia{}(\alphas\ {\small Down}), respectively.
The selected \alphas\ values are found to produce variations that are similar to the uncertainty 
band of cross-section measurements for \ttbar{} events with additional jet activity, as 
described in ref.~\cite{ATLAS:2012al}.
The effect of colour reconnection is estimated by generating a \powheg{}+\pythia{} sample in which no colour reconnection is allowed within \pythia{}, using the noCR Perugia 2011 tune~\cite{Skands:2010ak}.

The total inclusive
\ttbar\ cross-section for $pp$ collisions at $\sqrt{s} = 7 \tev$ is calculated to be $\sigma_{t\bar{t}}= 177^{+10}_{-11}$~pb for
a top-quark mass $m_{t} = 172.5$~\GeV. This
calculation is carried out at next-to-next-to-leading order (NNLO) in QCD including resummation of next-to-next-to-leading logarithmic (NNLL) soft gluon 
terms~\cite{Cacciari:2011hy,Beneke:2011mq, Czakon:2013goa,Czakon:2012pz, Czakon:2012zr,Baernreuther:2012ws} with \toppp{} (v2.0)~\cite{Czakon:2011xx}. 
All \ttbar~MC samples were generated with $m_{t} = 172.5$~\GeV\ and were normalised to the theoretical NNLO+NNLL cross-section prediction.

For the simulation of background processes, samples of $W$ and $Z$ bosons 
produced in association with jets were generated using  \alpgen{} (v2.13)
with LO matrix elements for up to five inclusive associated partons.  
The \cteq{} PDF set and the \herwig{} parton shower were used.
In addition to the inclusive jet-flavour processes, separate samples of $Wb\bar{b}\!+\!\mathrm{jets}$, $Wc\bar{c}\!+\!\mathrm{jets}$, $Wc\!+\!\mathrm{jets}$ and $Zb\bar{b}\!+\!\mathrm{jets}$
matrix-element processes 
with three additional partons
were generated and the overlap between them removed.
The normalisation of the \wjets\ samples were determined from data as described in section~\ref{sec:background-estimation}.
The \zjets\ samples are normalised to the cross-section obtained from an NLO QCD
calculation with \mcfm{} \cite{Campbell:1999ah} using the \mstw{} PDF set \cite{Martin:2009iq}.

A sample of $t$-channel single top-quark decays was generated using the \acermc{} generator~\cite{KER-0401} (v3.8), 
while \mcnlo{} was used to generate $Wt$-channel and $s$-channel processes.  Each of these samples is
normalised according to an NLO+NNLL calculation for the corresponding  
$t$-channel~\cite{singletop_tchan}, $s$-channel~\cite{singletop_schan} 
and $Wt$-channel~\cite{singletop_Wtchan} processes.  Diboson events ($WW$, $WZ$, $ZZ$) were produced 
using \herwig{} and normalised to the cross-section obtained from an NLO QCD calculation 
with \mcfm{} using the \mstw{} PDF set.

To properly simulate the LHC environment, additional inelastic $pp$ interactions were generated with
 \pythia{} using the AMBT1 tune \cite{ATLAS-CONF-2010-031}
and then overlaid on
the hard process.
The MC events are weighted such that the distribution of the generated mean number of $pp$ 
collisions per bunch crossing ($\langle \mu \rangle$) matched that of the data-taking period.
The particles from additional interactions are added before the 
signal digitisation and reconstruction steps of the detector simulation,
but are not used within the particle-level measurement defined 
in section~\ref{sec:meas-def}.

The response of the detector to the generated events is determined by a full \geant{}~\cite{AGO-0301} simulation of the ATLAS
detector~\cite{:2010wqa}. This is performed for all samples except for the ISR/FSR variations, colour reconnection 
and \powheg{}+\herwig{} MC samples. For those samples a faster simulation which parameterises the ATLAS calorimeter response is used instead~\cite{:2010wqa}. 

\section{Data sample and event selection}
\label{sec:data_event_sel}

The data are selected from the full 2011 data-taking period. Events are required to meet baseline data quality criteria during stable LHC running periods. These criteria reject data with significant detector noise or read-out problems and depend on the trigger conditions and the reconstruction of physics objects. 
The resulting data set corresponds to an integrated luminosity of $4.59\pm0.08$~\ifb~\cite{Aad:2013ucp}.  
During this period, the LHC delivered instantaneous luminosities that were sufficient to produce several $pp$ interactions within the same bunch crossing (in-time pile-up).
Interactions in adjacent bunch crossings also influenced the detector and readout signals in the selected bunch crossing.  The mean number of in-time pile-up interactions, $\langle \mu \rangle$, was measured by averaging over all $pp$ bunch crossings in a given luminosity block.  The
average value of $\langle \mu \rangle$ was approximately 5 at the beginning of the data-taking period and as high as 18 by the end of the 2011 run.  

\subsection{Object reconstruction}
\label{object-selection}

Primary vertices are formed from tracks reconstructed in the ID.  
The selected primary vertex is required to include at least four reconstructed tracks satisfying $\pT> 0.4$~GeV and to be consistent with the $pp$ beam collision region in the transverse plane.  
In the cases where more than one primary vertex with at least four tracks is 
reconstructed, the vertex with the highest $\sum \pT^2$ 
of the associated tracks
is chosen and assumed to be associated with the hard process.

Electron candidates are identified as electromagnetic energy deposits (clusters) matched to a reconstructed track in the 
ID~\cite{Aad:2014fxa}. Selected electrons are required to satisfy 
stringent identification criteria. The reconstructed tracks are required to have 
a minimal number of pixel and SCT hits among those expected along the electron trajectory
and at least a minimum number of high-threshold TRT hits. 
The longitudinal and lateral shower profiles in the calorimeter are required to match those expected 
for an electron, with a satisfactory match between the cluster energy and the reconstructed track momentum. 

To reduce the rate of events with non-prompt and fake lepton signatures from multi-jet background processes,
electrons are required to be isolated within both the calorimeter and ID.  The calorimeter isolation is defined using a cone of size
$\Delta R = 0.2$ around the electron direction.  The transverse energy sum of the clusters found in the cone is calculated and required
to be less than 10\% of the electron transverse energy, after excluding the calorimeter cells associated with the electron cluster and correcting for leakage from the electron cluster.  
The ID-based isolation
is calculated using the summed track-\pT\ within a $\Delta R =0.3$ cone around the electron direction and is required to be less 
than 10\% of the electron track \pT.  
Electrons are selected by requiring $\pT>25$~\GeV{}
in the range $|\eta|<2.47$, excluding the barrel/end-cap transition region of $1.37<|\eta|<1.52$.  
Electrons with $\pT>15$~\GeV\ are used for 
the object overlap removal discussed later in this section and to remove events with two or more leptons 
as discussed in section~\ref{sec:event-selection}.

Muon candidates are required to be composed of a reconstructed track in the MS combined with a track in the ID~\cite{Aad:2014zya}.  
Track quality criteria are used to reduce the non-prompt and fake muon background from multi-jet processes 
and to select a sample with improved \pT\ resolution.  
Reconstructed tracks are required to have a hit in the innermost pixel layer if expected from the track trajectory and at least a minimum number of pixel and SCT hits, 
set below the number of expected hits on a muon trajectory.
Muons crossing the TRT are required to have a hit pattern consistent with a well-reconstructed track.

To further reduce the fake muon 
background, muons are required to be isolated within the calorimeter and ID.  The calorimeter isolation is determined using 
using calorimeter energy deposits in
a  $\Delta R = 0.2$ cone around the direction of the muon and is required to be less than 4~\GeV.  
The ID isolation is determined from the summed \pT\ of tracks in a $\Delta R = 0.3$  
cone around the direction of the muon and is required to be less than 2.5~\GeV,
excluding the \pT\ of the muon. The muon channel event selection requires the reconstruction of one muon with $\pT > 25$~\GeV\, associated
with the selected primary vertex. Muons with $\pT>15$~\GeV\ are
used to define an additional lepton veto discussed in section~\ref{sec:event-selection}.  Both types of muons are selected within $|\eta|<2.5$.

Topological clusters~\cite{Lampl:1099735} are formed from calorimeter energy deposits.   
Jets are reconstructed from these clusters 
with the anti-$k_t$ algorithm~\cite{Cacciari:2008gp} 
with a radius parameter of 0.4.  The jets are calibrated using the EM+JES scheme,
where the jet energy scale (JES) is derived as a correction of the initial calorimeter calibration set for electromagnetic showers
\cite{Aad:2011he}. The jet energy is corrected for the effect of additional $pp$ collisions in data and MC events. 

To correct for energy losses due to non-compensation in the calorimeter,
uninstrumented material and detector subsystems in front of the calorimeter,
jet energy corrections factors are applied that depend on the jet energy and the jet $\eta$
to achieve a calibration that matches the energy of stable particle jets in simulated events
(excluding neutrinos and muons).

Differences between data and Monte Carlo simulation are evaluated using {\it in situ} techniques and are corrected  
in an additional step \cite{Aad:2014bia}.  The {\it in situ} calibration exploits the \pt{} balance 
in events with a $Z$ boson ($Z$+jet) or a photon ($\gamma$+jet) and a recoiling jet and dijet events.
Z+jet and $\gamma$+jet data are used to set the 
JES in the central detector region, while \pt balancing in dijet events is used to 
achieve an $\eta$ intercalibration of jets in the forward region with respect to central jets

The calibrated jets are required to satisfy $\pT > 25$~\GeV\ and be within the range $|\eta| < 2.5$.  Jets associated with large energy deposits from additional $pp$ interactions are removed by requiring that the \pT\ sum of the reconstructed tracks  matched with the jet and the selected primary vertex is at least 75\% of the total
 \pT\ sum of all tracks associated to the jet.  This quantity is referred to as the jet vertex fraction (JVF).  Jets satisfying $\pT > 50$~\GeV\ are always accepted and jets having no associated tracks are also accepted.

The MV1 algorithm~\cite{ATLAS-CONF-2011-102} is used to select jets associated with $b$-hadron decays.  The algorithm combines several tagging algorithms into a single neural-network-based discriminant.  Jets are identified as $b$-jets by using an MV1 discriminant tuned to achieve a 70\% tagging efficiency for jets with $\pT > 20$~\GeV\ in simulated \ttbar\ events.  The corresponding rejection factor for jets originating from gluons or light quarks is found to be approximately 130.

The \met\ and its associated azimuthal angle are reconstructed from the vector sum of the transverse momenta of the reconstructed objects (electrons, muons, jets) as well as the transverse energy deposited in calorimeter cells not associated with these objects, within the range $|\eta| < 4.9$.  The object classification scheme for the electrons, muons and jets used to 
calculate \met\ is chosen to be the same as the object definitions used in this analysis. 
Calorimeter cells not associated with an object are calibrated at the electromagnetic scale before being 
included in the \met\ calculation.  
This calibration scheme is similar to the one described in ref.~\cite{Aad:2012re}.

Electron and jet objects are reconstructed using separate algorithms that are run independently.  Jets are reconstructed from topological clusters as described above, with no distinction made between identified electron and jet energy deposits within the electromagnetic and hadronic calorimeters. Jets associated with an electron energy deposit are discarded using angular matching.  For each electron, the jet with an axis closest to that of the electron direction, within 
$\Delta R < 0.2$, is discarded. To remove leptons from heavy-flavour decays, the lepton is discarded if the lepton is found to be within $\Delta R < 0.4$ of a selected jet axis.

\subsection{Event selection}
\label{sec:event-selection}

Data were collected by requiring either a high-\pT\ electron trigger, based on calorimeter energy deposits, shower shape
and track quality constraints, or a high-\pT\ muon trigger that
included a reconstructed track in the MS matched with a track in the ID.  The electron trigger \pt\ threshold was
either 20~\GeV\ or 22~\GeV, depending on the data-taking period,
whereas the muon trigger \pT\ threshold remained at 18~\GeV\ for the duration of the entire 2011 data taking period.

The selected events are required to contain at least one reconstructed primary vertex.  A small number of events are rejected that
 included one or more jets of $\pT > 20$~\GeV\ with energy
that is identified as being from noise in the calorimeter electronics, from non-$pp$ collision background sources or
cosmic-ray showers.  Events where an identified electron and muon share the same reconstructed track in the ID are also removed.

Events are classified in the electron (muon) channel by the presence of one electron (muon) with $\pT > 25$~\GeV, no additional
electron or muon with $\pT > 15$~\GeV\ and at least four reconstructed jets with $\pT > 25$~\GeV\ and $|\eta|<2.5$ where at
least two are identified as $b$-jets. To reduce the background contribution from non-prompt or fake leptons, 
$\met > 30$~\GeV\ and $\mtw > 35$~\GeV\ are also required.  To reduce the effects of jet merging and migrations 
within the \pT\ ordering of the jets described in section~\ref{sec:meas-def}, 
events with a pair of reconstructed jets separated by $\Delta R < 0.5$ are vetoed.

\subsection{Estimation of backgrounds}
\label{sec:background-estimation}

The dilepton \ttbar\ final states constitute the most important background to this analysis, 
followed by single top-quark production, by $W$ boson production in association with jets 
including jets initiated by charm- and bottom-quarks (\wjets) and by
processes where only many jets are produced (multi-jet production).
In comparison, $Z$ boson production in association with jets (\zjets) 
and diboson production processes 
are smaller background components.

The fraction of dilepton events that remain after applying the full event selection is
evaluated with a MC simulation and removed from the data. 
A bin-by-bin correction factor derived from the baseline \powheg{}+\pythia{} \ttbar{} Monte Carlo simulation
is used (see section~\ref{sec:corr_chancomb}).  
Contributions from single top-quark, \zjets\ and diboson ($WW$, $WZ$, $ZZ$) production
are evaluated using corresponding MC samples and theoretical
cross-sections for these processes, as discussed in section~\ref{sec:monte-carlo-samples}.

The overall normalisation of the \wjets\ MC sample was determined in the data via a lepton charge asymmetry 
measurement described in ref.~\cite{Aad:2014zka}.  
This method exploits the fact that the production of $W$ bosons at the LHC is charge-asymmetric
and also that the ratio of the number of $W^-$ to $W^+$ bosons
is more precisely known than the total number of $W$ bosons \cite{Kom:2010mv}.
Most of the other background processes result in lepton charge distributions that are symmetric.  
The numbers of events with positively and negatively charged leptons are measured in the data and are referred to 
as $N_\ell^{+}$ and $N_\ell^{-}$, respectively.  A MC simulation was used to estimate the
charge-asymmetric background from single top-quarks and to subtract 
that from the values of  $N_\ell^{+}$ and $N_\ell^{-}$.  
The number of \wjets\ events was then extracted from:
\begin{equation}
N_{W^+}+N_{W^-}=\frac{r_\mathrm{MC}+1}{r_\mathrm{MC}-1}(N_\ell^{+} - N_\ell^{-})
\nonumber
\end{equation}
where 
$r_\mathrm{MC}$ is 
the ratio of the \wpjets{} and \wnjets{} production cross-sections 
determined using the \wjets{} MC simulation
for the signal region kinematic cuts, and $N_{W^+}$ ($N_{W^-}$) is the number of $W^+$ ($W^-$) events.  
The \wjets\ normalisation was determined using the event selection of this analysis, 
but without the $b$-tagging requirement.  
The values of ($N_{W^+}+N_{W^-}$) are independently determined for $W\!+\!4$-jet and $W\!+\!\ge 5$-jet events
and are applied as inclusive scale factors to the \alpgen{} $W\!+$jet processes.
The normalisation was separately obtained for each of the MC systematic uncertainty evaluations
 listed in section~\ref{sec:systematic_uncertainties}.

The normalisation of the heavy-flavour fractions within the \wjets\ sample was determined by measuring
the number of $W\!+\!2$-jet events, without a $b$-tagging requirement and with the requirement of at
least one $b$-tag.  To make this measurement, the charge asymmetry technique described above was 
applied to both sets of selected events.  The number of events that have one or more $b$-tags is 
related to the number of events before the $b$-tagging 
requirement, the $b$-tagging probability and the flavour fractions in the \wjets\ sample. The simulated 
fractions for the heavy-flavour processes
($Wb\bar{b}\!+\!\mathrm{jets}$, $Wc\bar{c}\!+\!\mathrm{jets}$ and $Wc\!+\!\mathrm{jets}$) 
are determined with the ratio  
$Wc\bar{c}/Wb\bar{b}$ taken from simulation. The measurement 
used the number of events after requiring at least one $b$-tag where the 
overall normalisation of \wjets\ events was fixed using the values previously 
determined by the charge asymmetry method.  
The heavy-flavour fractions are then extrapolated 
from the $W\!+\!2$-jet selection to 
$W\!+\!\ge 4$-jet events using
the heavy-flavour fractions from the MC simulation.
As a result the $Wb\bar{b}\!+\!\mathrm{jets}$ process is scaled up by about 30\% 
with respect to the Monte Carlo prediction
with a relative systematic uncertainty of 20\% and the $Wc\!+\!\mathrm{jets}$ process is scaled down 
by about 15\% with a relative systematic uncertainty of 40\%. 
The scale factors are determined separately for the electron and muon channel.

Multi-jet production processes have a large cross-section and can  
provide a non-prompt or fake lepton signature
due to interactions with detector material, 
electromagnetic shower fluctuations and heavy-flavour decays.  
In the electron channel, jets and electrons from photon conversions or heavy-flavour decays 
can mimic isolated electrons from $W$ bosons.
In the muon channel, the background is dominated by the decay of heavy-flavour hadrons to muons.  

A matrix method~\cite{Aad:2010ey} is used to estimate the
number of background events using a second event sample for which the
lepton identification criteria are relaxed and the isolation requirements
removed (loose selection). The number of background events that pass the standard 
tight lepton selection described in section~\ref{sec:event-selection} ($N^{\rm tight}_{\rm fake}$) 
is then given by:
\begin{equation}
N^{\rm tight}_{\rm fake} = \frac{\epsilon_{\rm fake}}{\epsilon_{\rm real} - \epsilon_{\rm fake}}
(\epsilon_{\rm real} N^{\rm loose}  - N^{\rm tight}),
\nonumber
\end{equation}
where $\epsilon_{\rm real}$ and $\epsilon_{\rm fake}$ are the fractions of real and fake
leptons that pass the loose and the tight selection and  
$N^{\rm loose}$ ($N^{\rm tight}$) 
is the number of events with a lepton passing the loose (tight) selection.
The efficiency for a real lepton to pass the tight selection ($\epsilon_{\rm real}$)
is measured using a tag-and-probe technique using leptons from $Z$ boson decays.
The efficiency for a loose background event to pass the tight selection
($\epsilon_{\rm fake}$) is estimated in control regions dominated by background.
Contributions from \wjets{} and \zjets{} production are subtracted in
the control regions using simulation.

In the electron channel the control region to determine $\epsilon_{\rm fake}$ 
is defined by $\met < 20$~\GeV.
In the muon channel two control regions with similar numbers of events are used. 
One control region has events with low-\mtw\ while the other control region contains 
events where the selected muons have a large impact parameter. 
The efficiency $\epsilon_{\rm fake}$ 
is extracted separately from the two background-enriched samples and the average is used.

\section{Systematic uncertainties}
\label{sec:systematic_uncertainties}

The systematic uncertainties 
due to detector effects and the modelling of signal and background
are determined for each bin of the measured observables. Each systematic uncertainty is
evaluated by varying the relevant source by one standard deviation about its nominal value. This effect
is propagated through the event selection, unfolding and correction procedure. 
Deviations from the nominal case 
are evaluated separately for the upward and downward variations
for each bin, observable and channel. 
The total systematic uncertainty for each bin is calculated by adding the individual systematic 
contributions for that bin in quadrature.

The uncertainty on the \ttbar{} signal and background modelling are
discussed in sections~\ref{sec:systematic_uncertainties_signal_modelling}
and \ref{sec:systematic_uncertainties_background_modelling}. The detector-related
uncertainties are presented
in section~\ref{sec:systematic_uncertainties_experimental_uncertainties}. 

\subsection{Signal modelling}
\label{sec:systematic_uncertainties_signal_modelling}

The uncertainty due to the choice of MC generator to model \ttbar{} events 
is estimated by comparing
 \mcnlo{}+\herwig{} and \alpgen{}+\herwig, while the
colour reconnection modelling uncertainty is estimated 
by comparing the nominal \powheg+\pythia{} sample 
to the associated MC tune where the colour reconnection is disabled in \pythia{}.

The uncertainty in modelling the parton shower and hadronisation is evaluated 
from the relative difference between the \alpgen+\pythia{} \ttbar\ MC sample and the alternative \alpgen{}+\herwig{} sample. 

The evaluation of the uncertainty due to the choice of PDF set is obtained using the 
\nnpdf{}~\cite{Ball:2010de}, \mstw{}  and \cteqsixsix{}~\cite{Nadolsky:2008zw} PDF sets.  
An envelope of uncertainty bands is determined using the PDF4LHC recommendations~\cite{Botje:2011sn}. 

The uncertainty associated with the modelling of additional QCD radiation accompanying the \ttbar{} system is calculated 
by comparing the \alpgen{}+\pythia{} sample to 
the ones with varied radiation settings presented in 
section~\ref{sec:monte-carlo-samples}.  The variation
is achieved by changing the renormalisation scale associated with \alphas\ consistently in
the hard-scattering matrix element as well as in the parton shower.  The level of radiation through parton
showering~\cite{PUB-2013-005} is adjusted to encompass the ATLAS measurement
of additional jet activity in \ttbar\ events~\cite{ATLAS:2012al}. The uncertainty is estimated
as the maximum difference between the specialised samples and the 
nominal sample, with the uncertainty being symmetrised.

\subsection{Background modelling}
\label{sec:systematic_uncertainties_background_modelling}

The individual experimental and theoretical
uncertainties are used to calculate the uncertainty on the size of background 
contributions determined by MC simulation. 
This results in an uncertainty on the background subtraction
for all backgrounds except for those from \wjets{} processes and from multi-jet processes resulting in a non-prompt or fake lepton signature.

In the muon channel the normalisation and shape uncertainties are determined as the difference 
between the two multi-jet estimates described in section~\ref{sec:background-estimation}.
The normalisation uncertainty is determined to be $20$\%. 
The shape uncertainty is evaluated by two different linear combinations of the
two multi-jet estimates.
In the electron channel the normalisation uncertainty
is determined from the variation of the background estimate when the 
efficiencies for real and fake leptons are varied within their uncertainties. The uncertainty of the 
efficiency for real leptons is estimated by varying the fit parameters in the tag-and-probe method using $Z$ boson events.
The uncertainty on the efficiency for fake leptons 
is estimated by varying the \met\ cut between $15$~\GeV{} and $25$~\GeV{} 
and by relaxing the $b$-tag requirement to at least one $b$-tag. From these variations a normalisation uncertainty of $50$\% is assigned.

For the background contribution due to \wjets, the overall uncertainty from 
 the charge asymmetry normalisation method 
(see section~\ref{sec:background-estimation}) and the uncertainty on the flavour fractions are separately determined. 
A shape uncertainty is estimated by varying model parameters in the \wjets{} \alpgen{} simulation.
The total background uncertainty is evaluated by adding in quadrature each of the different background uncertainty contributions. 

\subsection{Experimental uncertainties}
\label{sec:systematic_uncertainties_experimental_uncertainties}

The experimental uncertainties refer to the quality of the detector simulation to describe the detector
response in data for each of the reconstructed objects.
These uncertainties affect the MC signal and background predictions, changing the numbers of events accepted.

The jet energy scale (JES) systematic uncertainty~\cite{Aad:2014bia}
is a major contributor to the overall systematic uncertainty
in all distributions affected by
the signal efficiency and bin migration. In the central region of the 
detector $(|\eta| < 1.7)$ it varies from 2.5\% to 8\% as a function of jet \pt{} 
and $\eta$ as estimated from {\it in situ} measurements of the detector response~\cite{Aad:2014bia}. It
incorporates uncertainties from the jet energy calibration, calorimeter response
to jets, detector simulation, and
the modelling of the fragmentation and underlying event, as well as other choices in the MC generation.
Additional sources of the JES uncertainty are also estimated. The main contributions are 
the intercalibration of the forward region detector response from the central regions of the detector, effects from the correction of 
additional $pp$ interactions, jet flavour composition, $b$-jet JES calibration and the presence of close-by jets.
Uncertainties due to different detector-simulation configurations used in the analysis and in the calibration are added as one
additional uncertainty parameter (``relative non-closure"). The JES uncertainty is evaluated using
a total of 21 individual components to model the uncertainty correlations as a function of the jet transverse momentum and the rapidity.  

The jet energy resolution (JER) has been found to be well modelled in simulation.
{\it In situ}  methods are used to measure the resolution, which MC simulation describes within 10\% for 
jets in the \pt\  range 30--500 \GeV{}~\cite{Aad:2012ag}. 
The jet reconstruction efficiency is also well modelled by the simulation and the uncertainty is evaluated by randomly removing 
simulated jets within the $1\sigma$ uncertainty of jet reconstruction efficiency measured in data~\cite{Aad:2014bia}. 

The uncertainties introduced by the JVF requirement 
used to suppress pile-up jets are estimated using events 
with a leptonic $Z$ boson decay and an associated high-$p_{\rm T}$ jet. 
The efficiency to select jets from the hard-scatter and the contamination
by jets produced by pile-up interactions are measured in appropriate control
regions and the agreement between data and MC simulation is evaluated.

The uncertainties related to the MC modelling of the lepton trigger, reconstruction and identification efficiency 
are evaluated by comparing high-purity events featuring leptons in data and simulation. These include
$Z \rightarrow ee$, $Z \rightarrow \mu\mu$ and $W \rightarrow e\nu$ events
in data and simulation, while \ttbar\ events are also included in the simulation studies~\cite{Aad:2014fxa}. Similar studies are
also performed for the lepton energy and momentum scales and resolutions. Since the two channels require different lepton flavours,
the electron uncertainty affects only the electron channel, and similarly for the muon channel. The electron uncertainty is approximately double the muon uncertainty. In both cases the uncertainty is small with little variation between bins.
The uncertainty on \met{} is determined by propagating all the uncertainties associated with the energy
scales and resolutions for leptons and jets to
the calculation of \met{}. 
Two additional uncertainties that are included originate from
the calorimeter cells not associated with any physics objects and the pile-up modelling. 
The systematic uncertainties associated with tagging jets originating from $b$-quarks are separated into three categories. These are the
efficiency of the tagging algorithm ($b$-quark tagging efficiency), the efficiency with which 
jets originating from $c$-quarks pass the $b$-tag requirement
($c$-quark tagging efficiency) and the rate 
at which light-flavour jets
are tagged (misidentified tagging efficiency). The efficiencies are estimated
from data and parameterised as a function of \pt\ and $\eta$ \cite{ATLAS-CONF-2011-102,ATLAS-CONF-2011-089}. 
The systematic uncertainties arise from factors used to correct the differences between simulation and data in each of the categories.
The uncertainties in the simulation modelling of the $b$-tagging performance are assessed by studying $b$-jets in
dileptonic \ttbar{} events \cite{ATLAS-CONF-2014-004}.  The
$b$-tagging efficiency is another major contributor to the overall systematic uncertainty 
and tends to slightly increase with \pt. The $c$-quark efficiency uncertainty is approximately constant at $\approx$ 2\% while the misidentification
uncertainty contributes at the percent level for all distributions.

The last experimental uncertainty evaluated is due to the measurement of the integrated luminosity.  This is dominated
by the accuracy of the beam separation scans and has an associated uncertainty of 1.8\%~\cite{Aad:2013ucp} that is assigned to each bin of the distributions and the MC background predictions.
With the exception of the MC and data statistics and the background modelling
all uncertainty components are correlated across the bins and for all observables.
\section{Reconstructed yields and distributions}
\label{sec:reco}

A summary of the number of selected data events, background contributions and total predictions is given in table~\ref{tab:yield}. 
Dilepton \ttbar\ events constitute the largest background followed by single top-quark production. 
The \wjets\ and non-prompt or fake lepton backgrounds are smaller in comparison.

\begin{table}[htbp]
\begin{center}
\begin{tabular}{ c|c|c }\hline \hline
  & \multicolumn{2}{c}{Yields} \\ \hline
Source                  & \ejets ($\pm \mathrm{stat.} \pm \mathrm{syst.}$) \ & \mujets\ ($\pm \mathrm{stat.} \pm \mathrm{syst.}$) \\ \hline
\ttbar\ (single-lepton)	& $7810  \pm 20 \pm 630$ 	                                & $\phantom{0}9410     \pm            22 \pm 700$ \\
\ttbar\ (dilepton)      & $\phantom{0}570 \pm \phantom{0}5 \pm \phantom{0}60$    	& $\phantom{00}670     \pm   \phantom{0}6 \pm \phantom{0}60 $\\
Single top-quarks       	& $\phantom{0}430 \pm \phantom{0}6 \pm \phantom{0}50$     & $\phantom{00}520     \pm   \phantom{0}7 \pm \phantom{0}50$ \\
\wjets\                 & $\phantom{0}260 \pm 15           \pm 110$	              & $\phantom{00}360     \pm            18 \pm 130$ \\
Non-prompt or fake leptons                & $\phantom{0}130 \pm 16           \pm \phantom{0}70$ 	  & $\phantom{000}56     \pm \phantom{0}4 \pm \phantom{0}11$ \\
\zjets\                 & $\phantom{00}52 \pm \phantom{0}4 \pm \phantom{0}13$    	& $\phantom{000}28     \pm \phantom{0}3 \pm \phantom{00}7$ \\
Diboson                 & $\phantom{000}7  \pm \phantom{0}1 \pm \phantom{00}1$     & $\phantom{0000}7   \pm \phantom{0}1 \pm \phantom{00}1$ \\
\hline
Expectation         & $9270 \pm 31 \pm 650$ & $11050 \pm 30 \pm 720$ \\ \hline
Observed            & $8791$    	& $10690$ \\
\hline \hline
\end{tabular}
\end{center}
\caption{The number of selected events from data, MC simulation and data-driven background estimates, in the electron and muon channels. The yields of
MC samples that are not constrained by fits to data shown are normalised to an integrated luminosity of 4.6 fb$^{-1}$. The uncertainties
 specific to the \ttbar\ MC sample (ISR, generator, fragmentation, etc.) are not included
in the \ttbar\ (single-lepton and dilepton) uncertainties given in this table.}
\label{tab:yield}
\end{table}

The reconstructed distributions of the observables of the $\pseudotopl$, $\pseudotoph$ and $\pseudotopl\pseudotoph$ system are shown in figures~\ref{fig:reco-pt-t}~to~\ref{fig:reco-m-tt}. This includes the $\pT$ and rapidity of the individual pseudo-top-quarks and the $\pT$, rapidity
and mass of the $\pseudotopl\pseudotoph$ system for the muon and electron channels. 

\begin{figure}[tbhp]
\centering
\subfigure[\label{fig:reco-pt-t-had-mu}]{\includegraphics[width=0.49\textwidth]{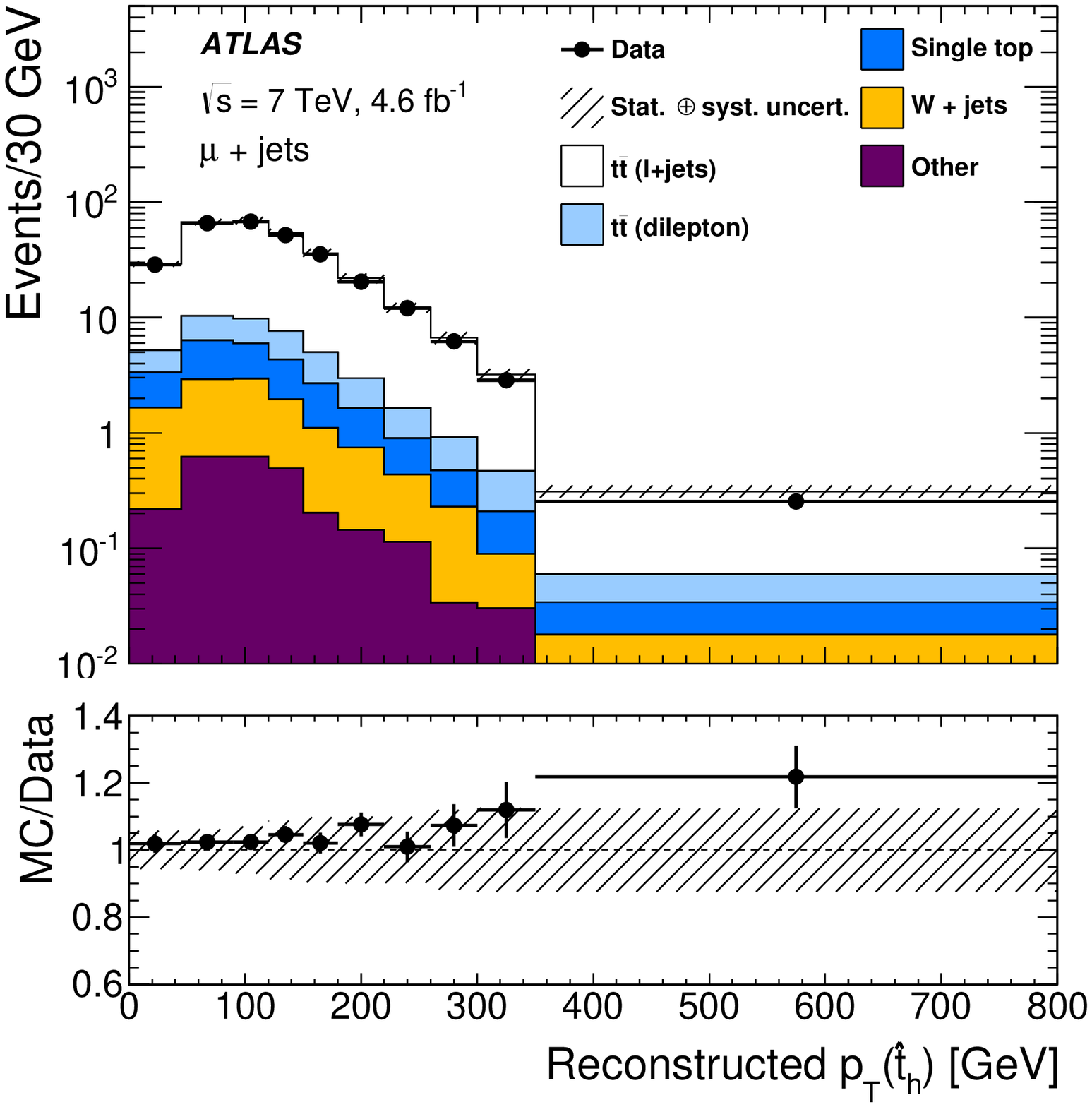}}
\subfigure[\label{fig:reco-pt-t-had-el}]{\includegraphics[width=0.49\textwidth]{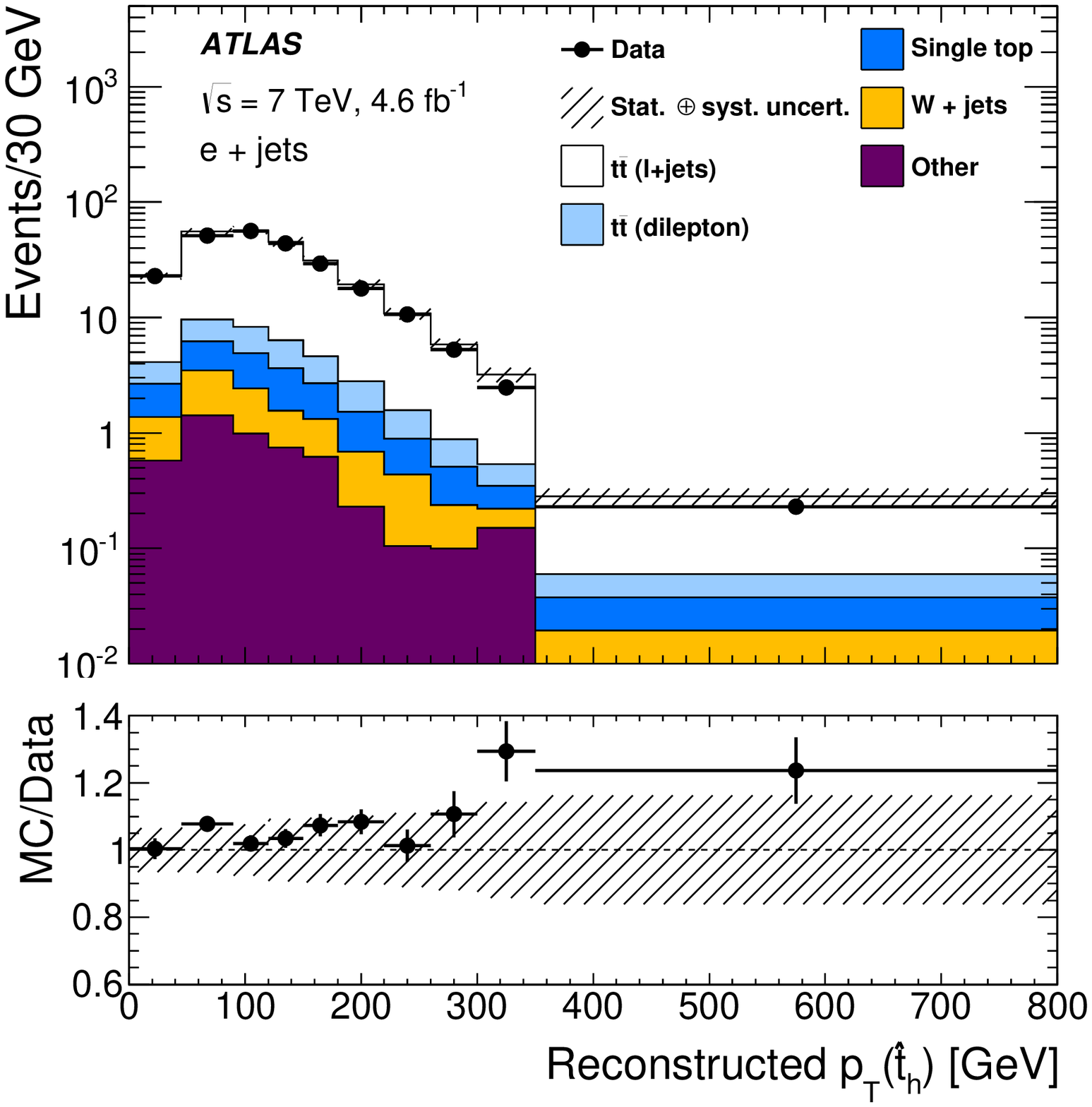}}
\subfigure[\label{fig:reco-pt-t-lep-mu}]{\includegraphics[width=0.49\textwidth]{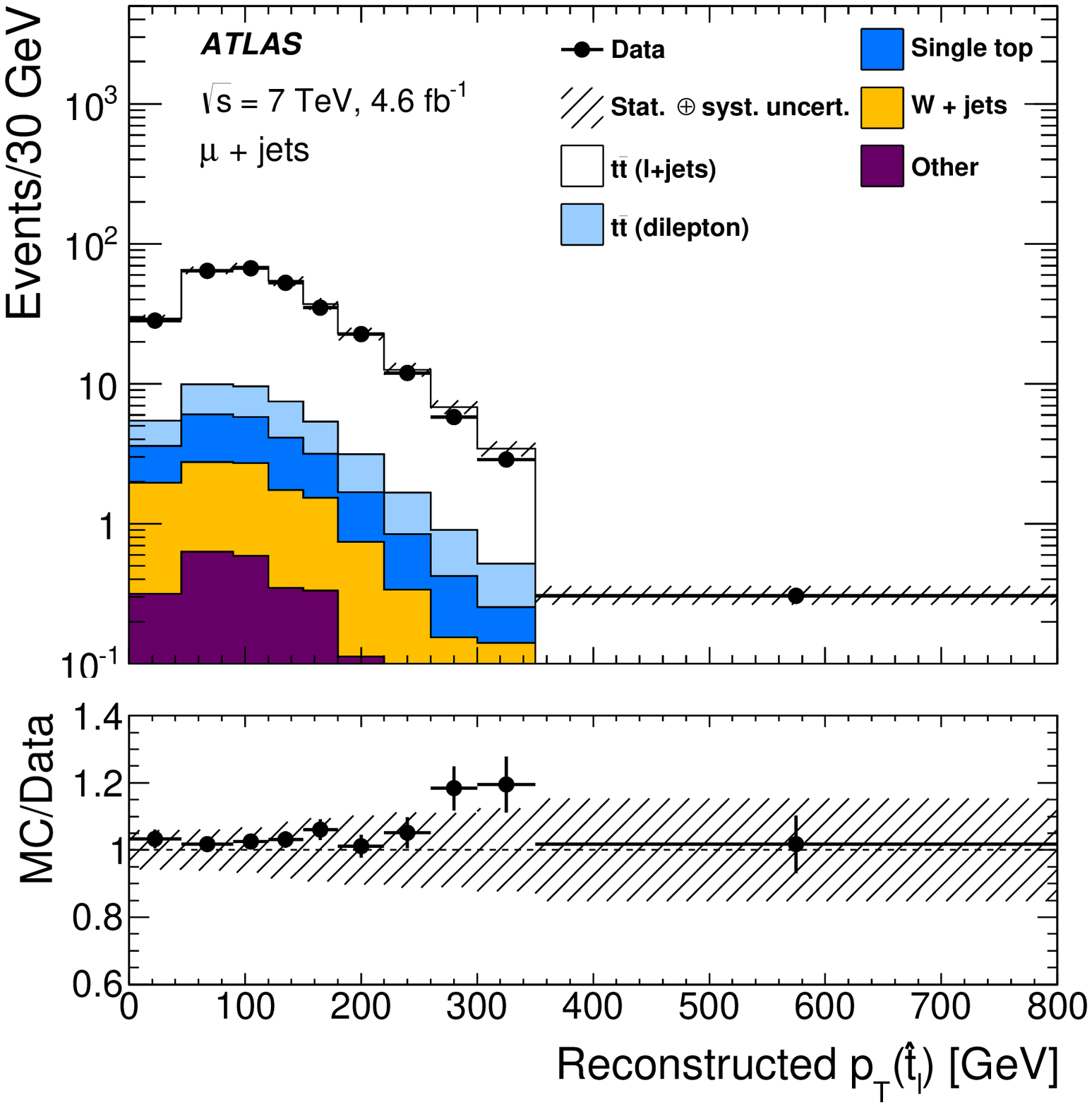}}
\subfigure[\label{fig:reco-pt-t-lep-el}]{\includegraphics[width=0.49\textwidth]{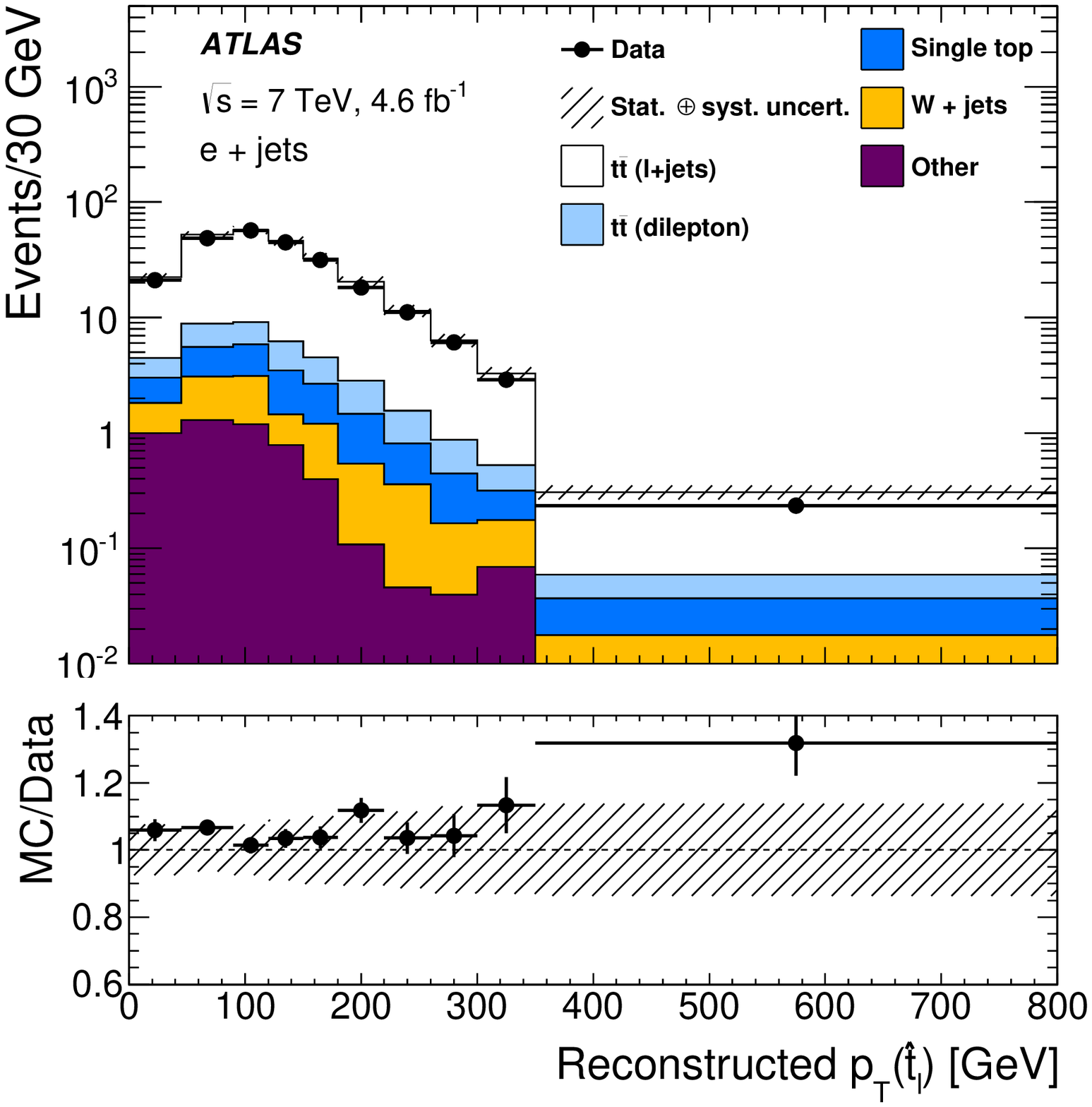}}
\caption{The reconstructed pseudo-top-quark $\pT$ in comparison to the MC signal and data-driven background models. The 
$\pTx(\pseudotoph)$ distributions: (a) the muon channel and (b) the electron channel. The $\pTx(\pseudotopl)$ distributions: (c) the muon channel and (d) the electron channel. 
 Signal and background processes are shown in different colours, with ``Other'' including the small backgrounds from diboson and  \zjets{} production, 
 and non-prompt or fake lepton signatures from multi-jet processes. The data are compared with predictions from background models and expected yields from simulated \ttbar\ events generated using \powheg{}+\pythia{} with the ``C'' variant of the Perugia 2011 tunes family.  
The error bars on the data points show the statistical uncertainty on the data, 
while the shaded band shows the total systematic and statistical uncertainty on the predicted yields.
The systematic uncertainty has a strong bin-to-bin correlation. }
\label{fig:reco-pt-t}
\end{figure}

\begin{figure}[tbhp]
\centering
\subfigure[\label{fig:reco-y-t-had-mu}]{\includegraphics[width=.49\textwidth]{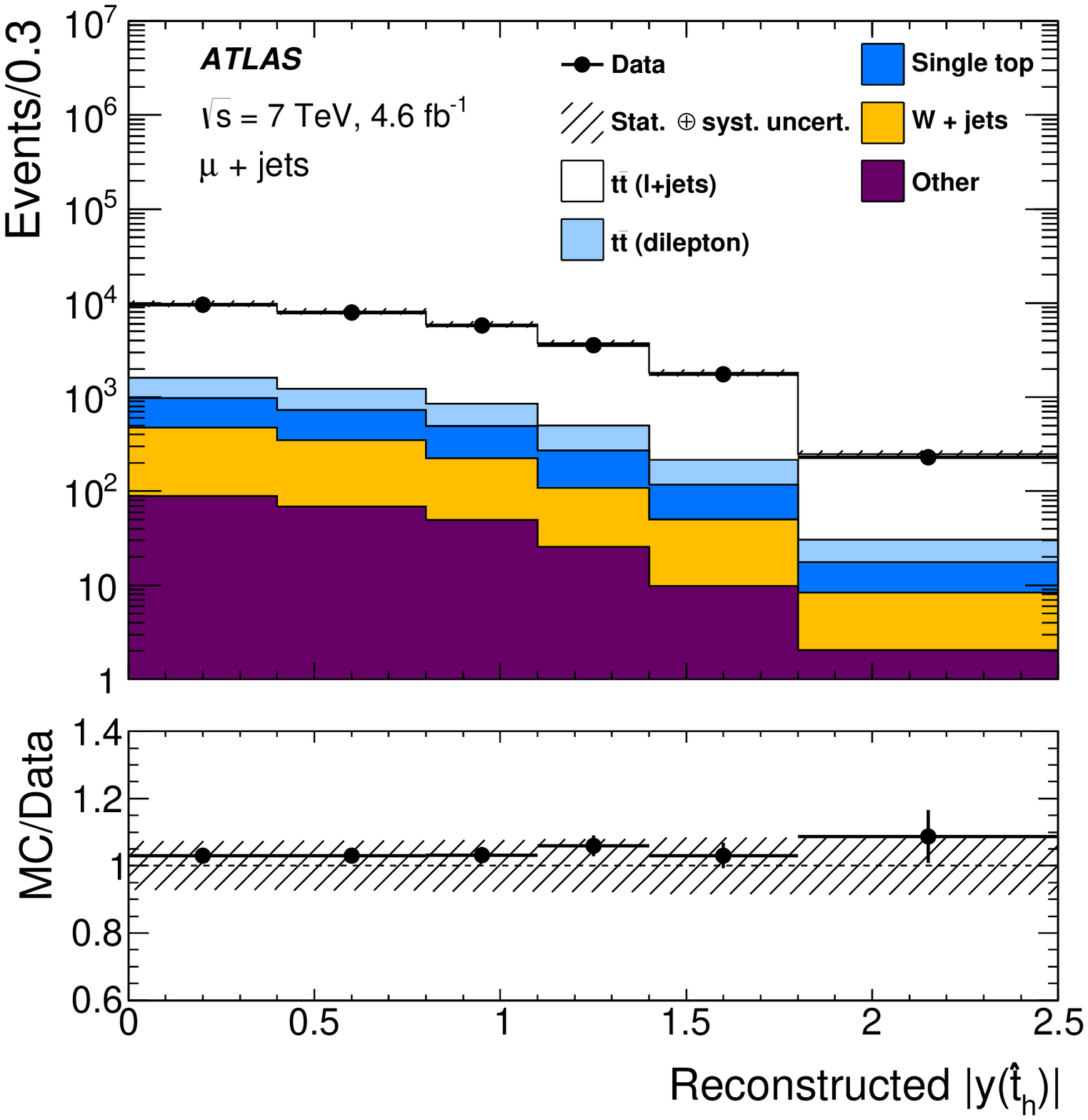}}
\subfigure[\label{fig:reco-y-t-had-el}]{\includegraphics[width=.49\textwidth]{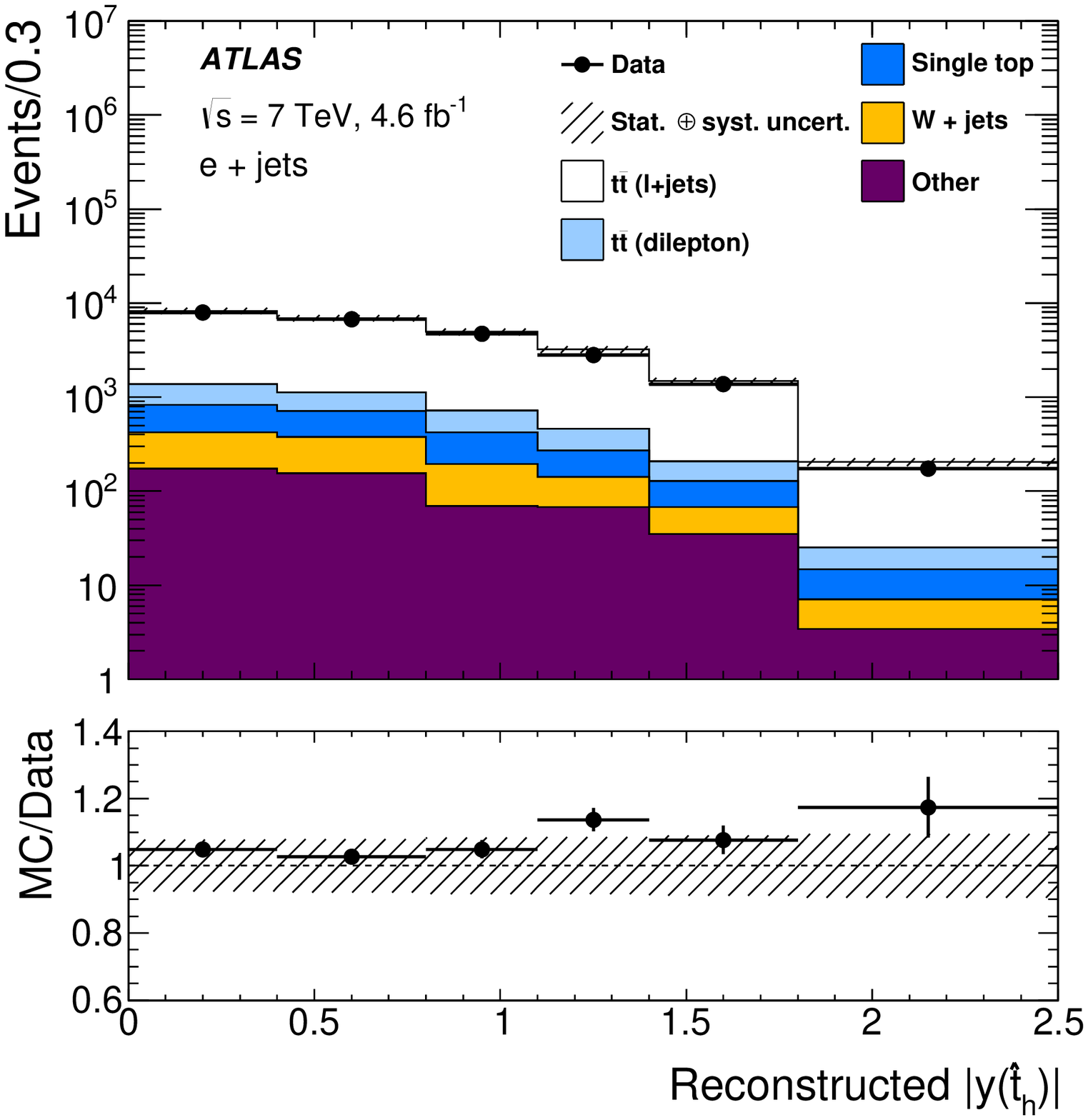}}
\subfigure[\label{fig:reco-y-t-lep-mu}]{\includegraphics[width=.49\textwidth]{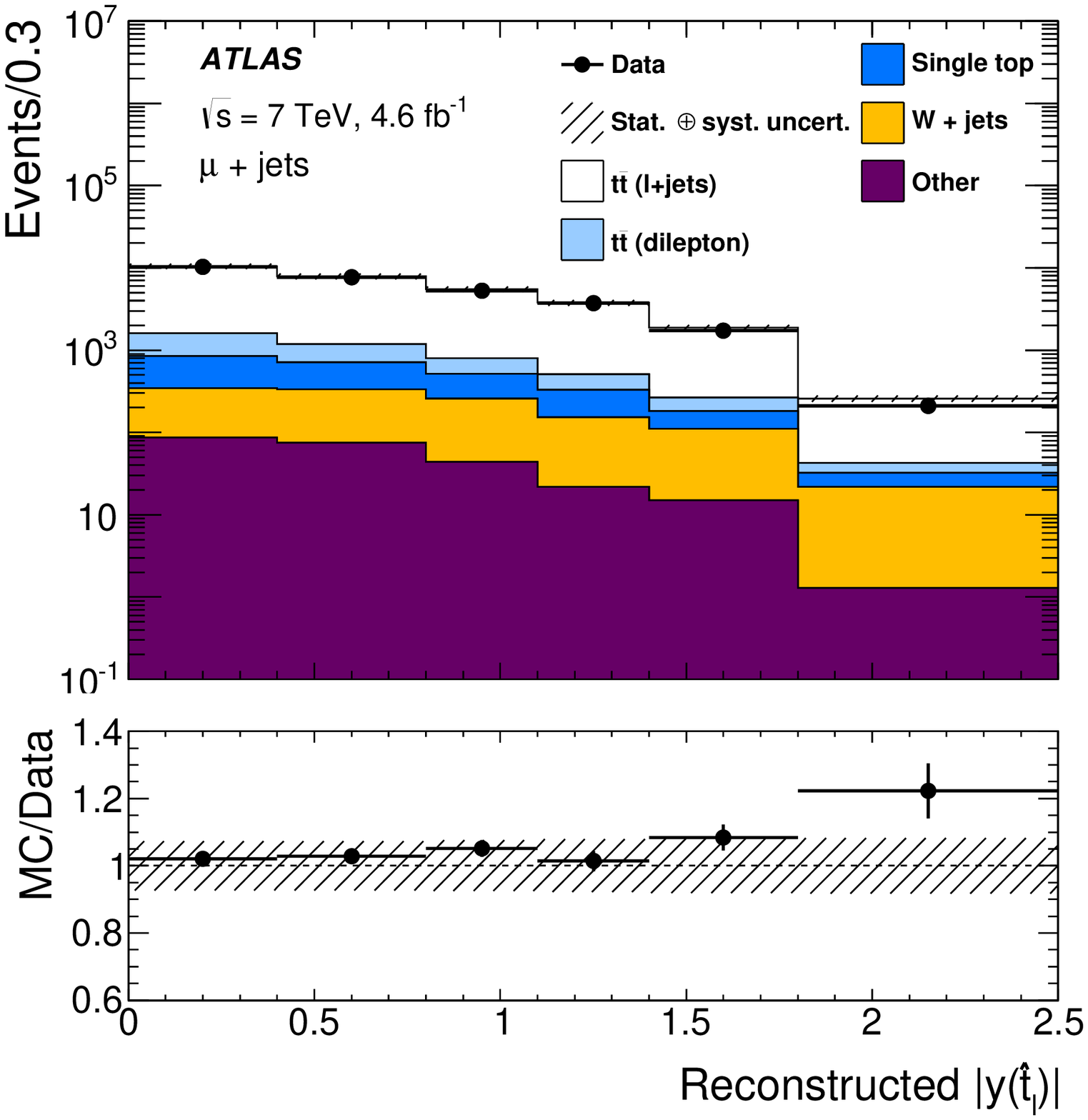}}
\subfigure[\label{fig:reco-y-t-lep-el}]{\includegraphics[width=.49\textwidth]{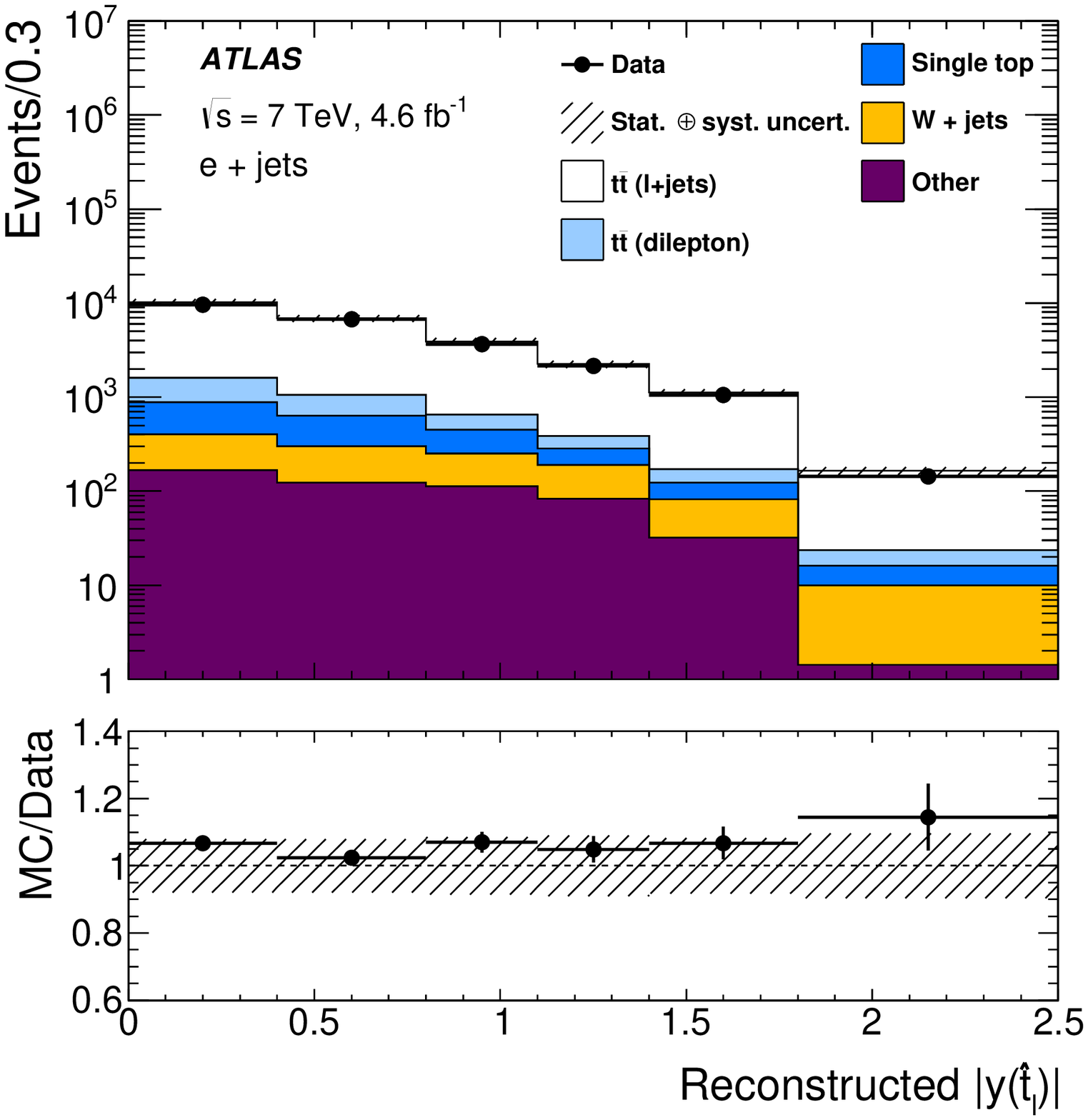}}
\caption{The reconstructed pseudo-top-quark rapidity in 
comparison to the MC signal and data-driven background models. The |$y(\pseudotoph)$| distributions:  (a) the muon channel and (b) the electron channel. The 
|$y(\pseudotopl)$| distributions:  (c) the muon channel and (d) the electron channel. The \powheg{}+\pythia{} MC generator with the Perugia 2011C tune is used for the \ttbar\ 
signal estimate. Signal and background processes are shown in different
colours, with ``Other'' including small backgrounds from diboson and \zjets{} production, as well as non-prompt and fake leptons from multi-jet processes.
The shaded band shows the total systematic and statistical uncertainties on the signal plus background expectation.}
\label{fig:reco-y-t}
\end{figure}

\begin{figure}[tbhp]
\centering

\subfigure[\label{fig:reco-pt-tt-mul}]{\includegraphics[width=.49\textwidth]{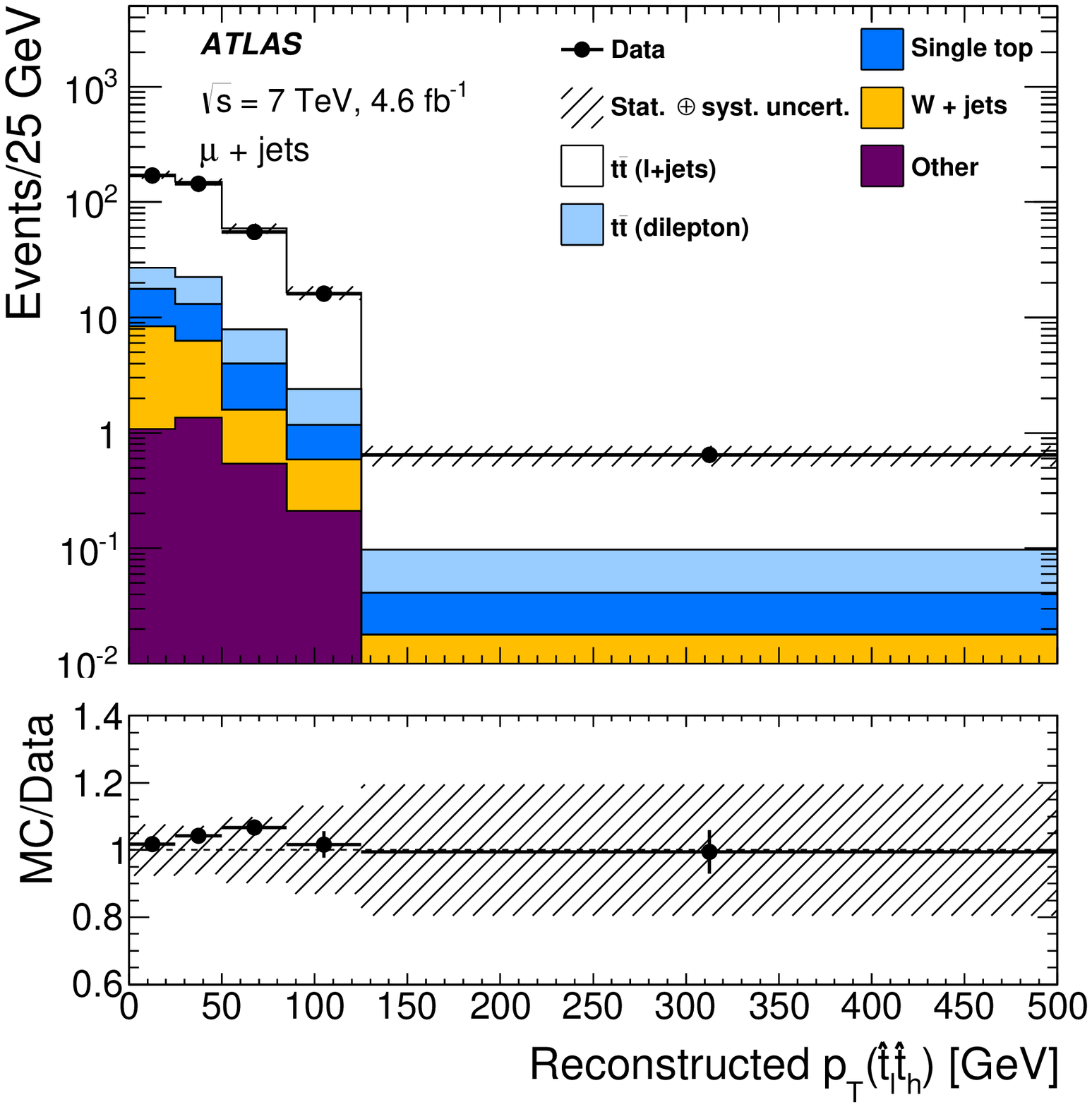}}
\subfigure[\label{fig:reco-pt-tt-el}]{\includegraphics[width=.49\textwidth]{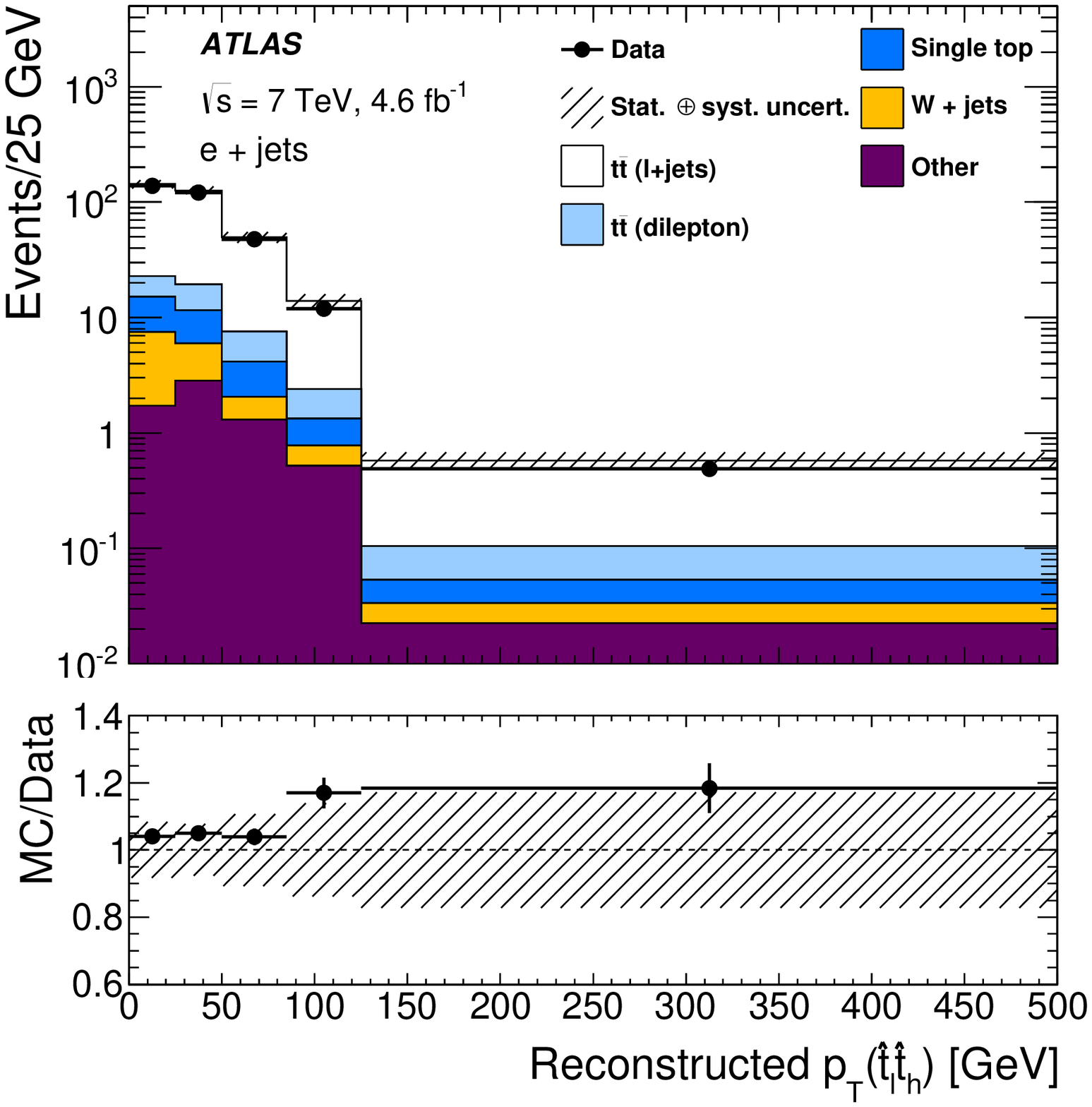}}

\caption{The reconstructed $\pT$ for the system of pseudo-top-quark pairs $\pseudotopl\pseudotoph$ in
comparison to the MC signal and data-driven background models. The $\pTx(\pseudotopl\pseudotoph)$ 
distributions: (a) the muon channel and (b) the electron channel. Signal
 and background processes are shown in different
colours, with ``Other'' including small backgrounds from diboson and \zjets{} production, as well as non-prompt and fake leptons from multi-jet processes. The \powheg{}+\pythia{}
 MC generator with the Perugia 2011C tune is used for the \ttbar\ signal estimate. The shaded band shows the total systematic and statistical 
uncertainties on the signal plus background expectation.}
\label{fig:reco-pt-tt}
\end{figure}

\begin{figure}[tbhp]
\centering

\subfigure[\label{fig:reco-y-tt-mu}]{\includegraphics[width=.49\textwidth]{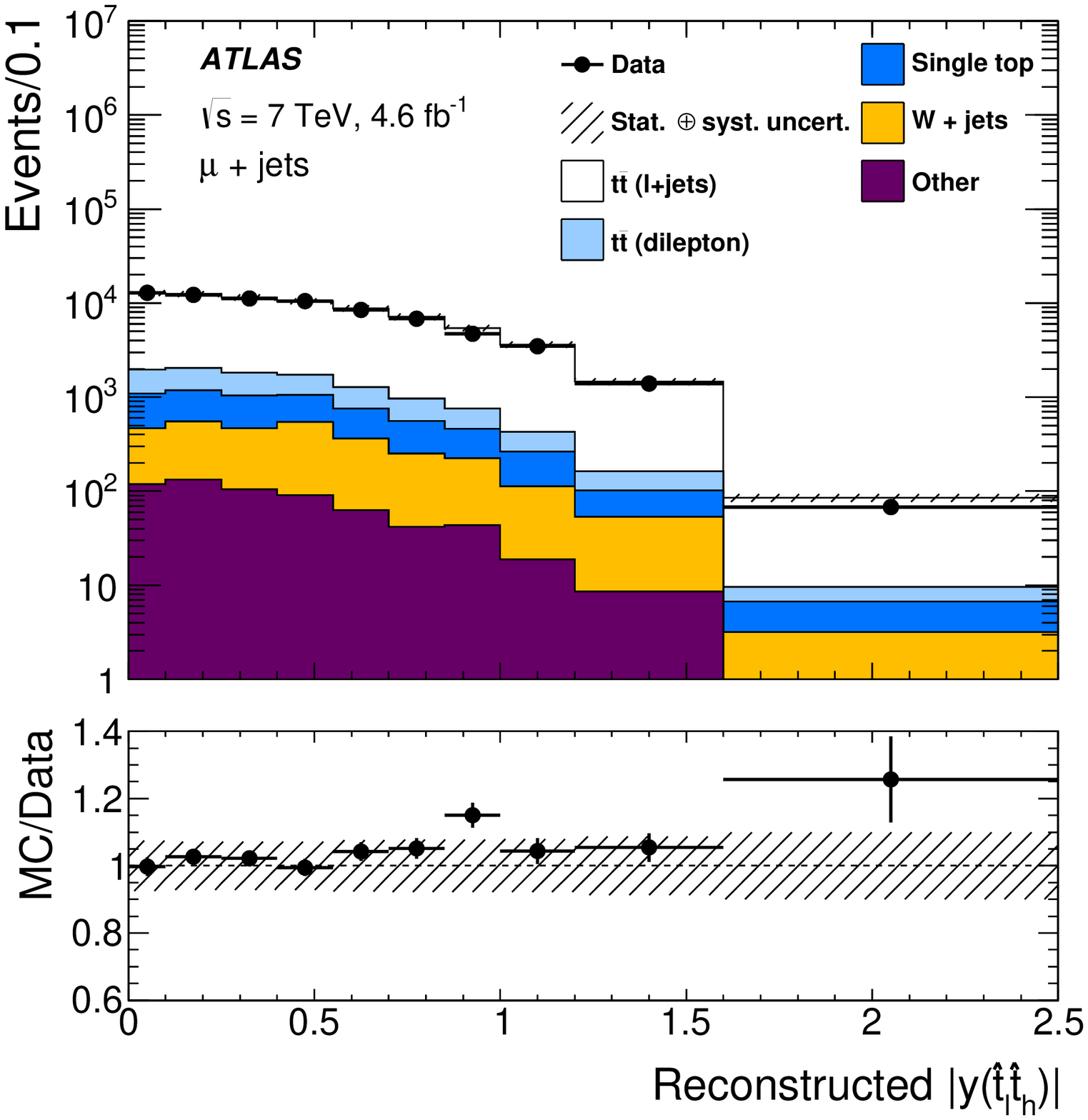}}
\subfigure[\label{fig:reco-y-tt-el}]{\includegraphics[width=.49\textwidth]{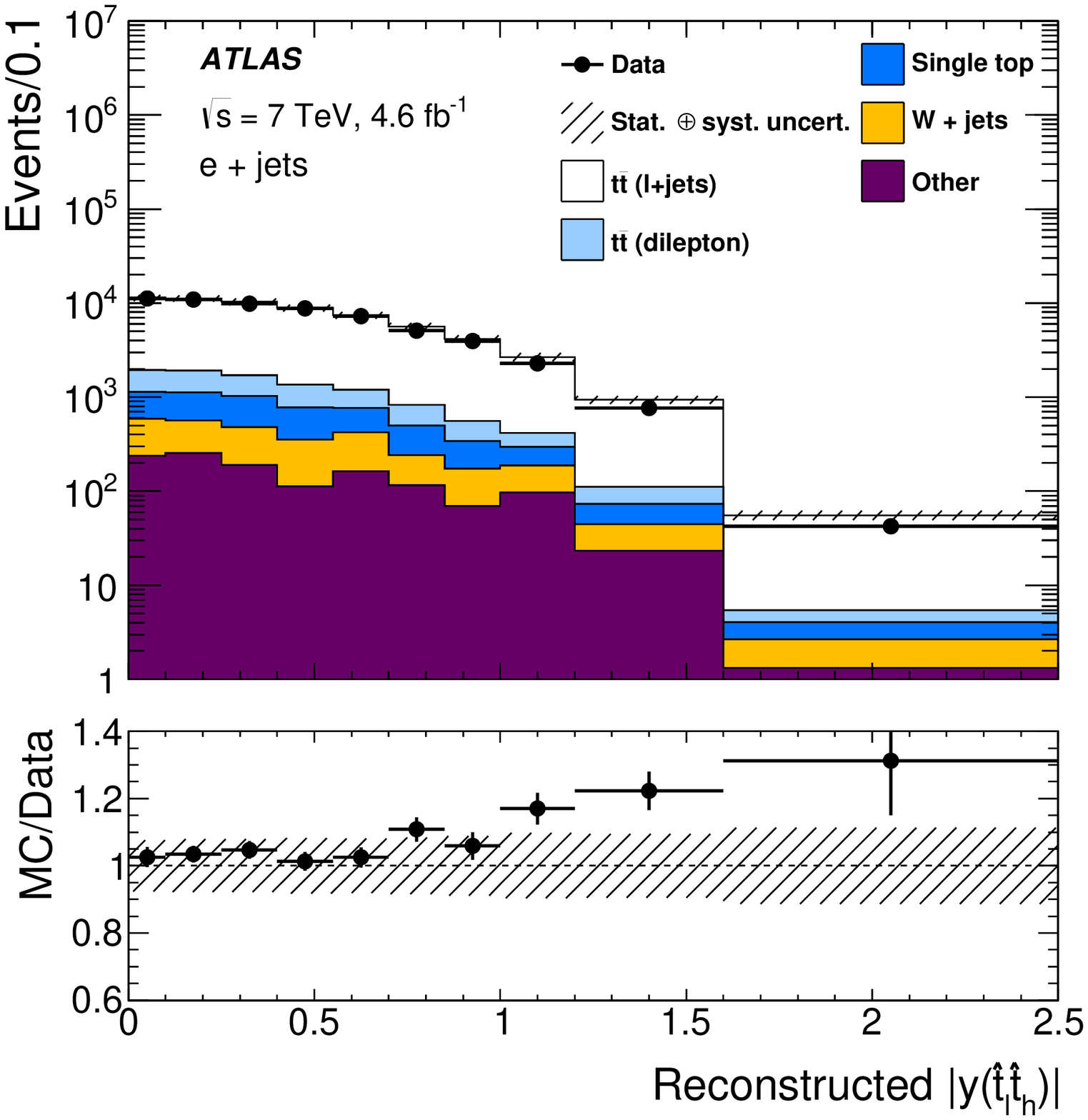}}

\caption{The reconstructed rapidity for the 
pseudo-top-quark system $\pseudotopl\pseudotoph$ 
in comparison to the MC signal and data-driven background models. The    
 $y(\pseudotopl\pseudotoph)$ distributions: (a) the muon channel and (b) the electron channel. Signal and background processes are shown in different
colours, with ``Other'' including small backgrounds from diboson and \zjets{} production, as well as non-prompt and fake leptons
 from multi-jet processes. The \powheg{}+\pythia{}
 MC generator with the Perugia 2011C tune is used for the \ttbar\ signal estimate. The shaded band shows the total systematic and statistical 
uncertainties on the signal plus background expectation.}
\label{fig:reco-y-tt}
\end{figure}

\begin{figure}[tbhp]
\centering
\subfigure[\label{fig:reco-m-tt-mu}]{\includegraphics[width=.49\textwidth]{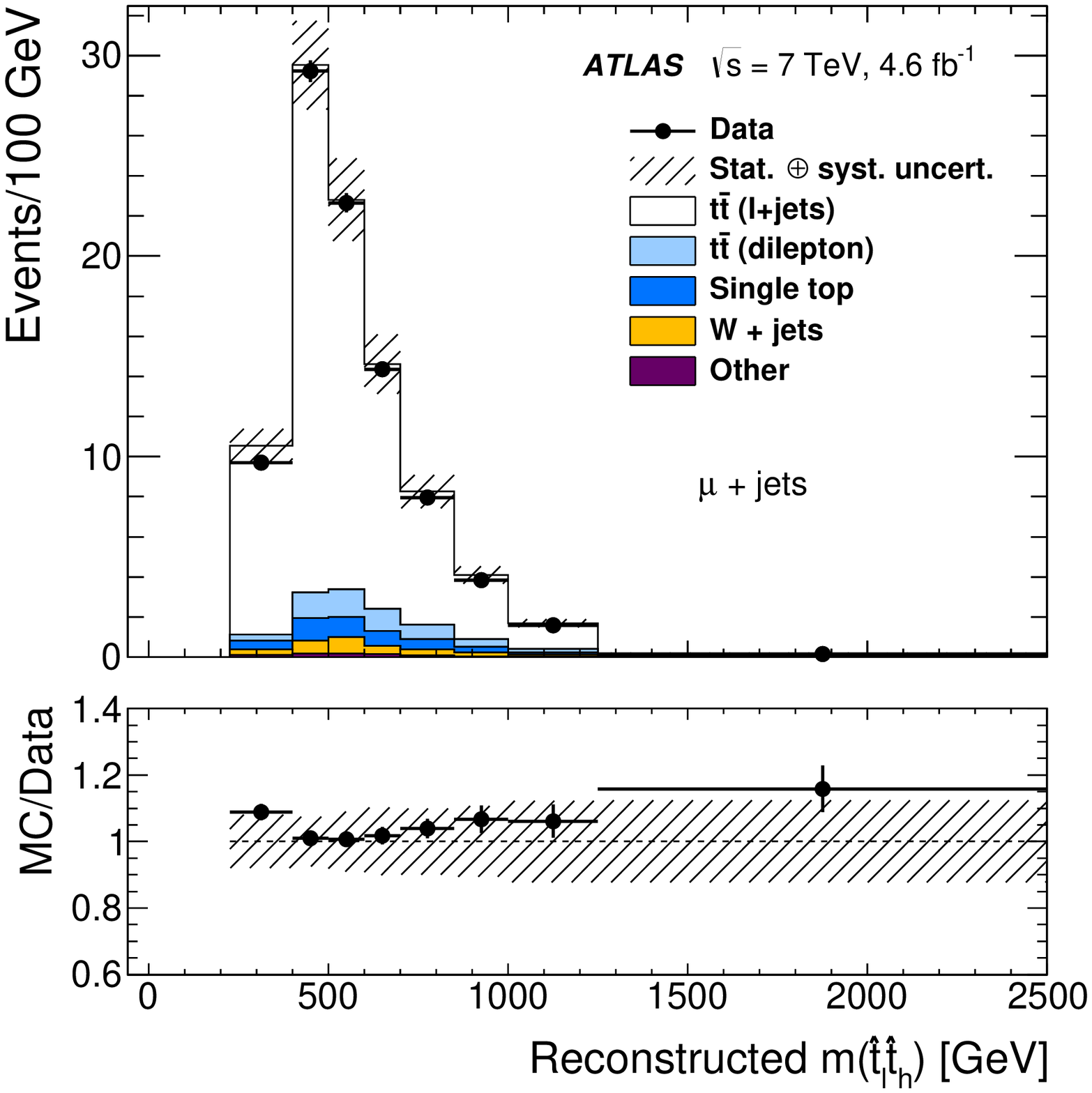}}
\subfigure[\label{fig:reco-m-tt-el}]{\includegraphics[width=.49\textwidth]{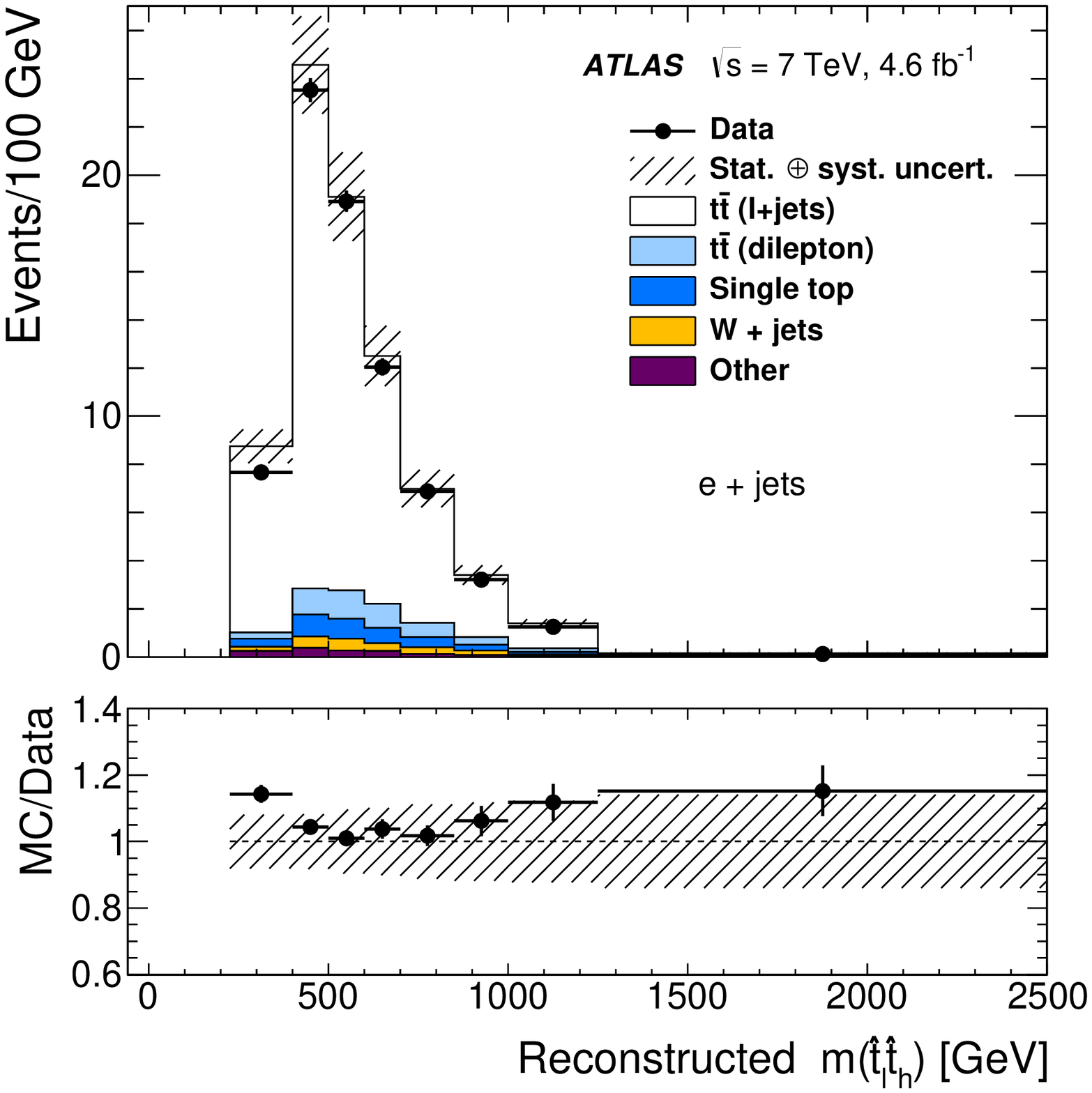}}
\caption{The reconstructed $m(\pseudotopl\pseudotoph)$ for the 
pseudo-top-quark system $\pseudotopl\pseudotoph$ 
in the muon (a) and electron (b)  channel. The distributions
are shown in comparison to the MC signal and data-driven background models. Signal and background processes are shown in different
colours, with ``Other'' including small backgrounds from diboson and \zjets{} production, as well as non-prompt and fake leptons from multi-jet
processes. The \powheg{}+\pythia{} MC generator with the Perugia 2011C tune is used for the \ttbar\ signal estimate.
The shaded band shows the total systematic and statistical uncertainties on the signal plus background expectation.}
\label{fig:reco-m-tt}
\end{figure}

In each region, the expected number of events agrees with the number observed in the data.
The data are shown using bin sizes that correspond to one standard deviation resolution in the hadronic variables, except 
in the tails of the distributions where the bin width is increased to reduce the statistical fluctuations.

\clearpage

\section{Corrections for detector effects and channel combinations}
\label{sec:corr_chancomb}

Each of the reconstructed pseudo-top-quark observable 
distributions is corrected for the effects of detector efficiencies and 
resolution. All distributions are
presented within the kinematic range defined in section~\ref{sec:meas-def}, which
is close to the acceptance of the reconstructed object and event selections, such that model dependencies from regions of phase space outside of the 
acceptance are minimised.  Section~\ref{sec:correction} describes the correction procedure, section~\ref{sec:propagation} describes the 
propagation of the statistical and systematic uncertainties to the final distributions, and section~\ref{sec:combination} describes the 
combination of the results obtained in the electron and muon channels.

\subsection{Correction procedure}
\label{sec:correction}

The reconstructed pseudo-top-quark observable
distributions are
corrected as follows:

\begin{equation}\label{eqn:corrections}
\Nparti= \fpartrecoi \cdot \sum_j \mrecopartij \cdot \fmisassignj \cdot \frecopartj \cdot (\Nrecoj - \Nbgndj)
\nonumber
\end{equation}

where \Nrecoj{} (\Nparti) is the number of reconstructed (fully corrected) events 
in a given reconstructed observable bin $j$ (particle-level observable bin $i$),
and \Nbgndj\ is the number of background events 
estimated as explained in section~\ref{sec:background-estimation}.
The correction factors  $\fpartrecoi$, $\fmisassignj$ and  $\frecopartj{}$ are discussed below.

Detector resolution effects on the reconstructed pseudo-top-quark observables
are corrected with an iterative Bayesian unfolding procedure~\cite{D'Agostini:1994zf}
using a response matrix \Mres\ that describes the migration 
between the detector-level observable bin $j$ and particle-level observable bin $i$.
To ensure a one-to-one relationship between the particle-level
and the detector-level observables, the response matrix is constructed from events
where for each detector-level pseudo-top-quark object a particle-level 
pseudo-top-quark object can be matched.
The matching is based on the angular differences between the components of the
pseudo-top-quarks. Detector-level and particle-level jets (${\rm jet}/{\rm jet}'$) 
and leptons ($\ell/\ell^\prime$) are considered matched if they 
satisfy $\Delta R({\rm jet},{\rm jet}') < 0.35$ and $\Delta R(\ell,\ell^\prime) < 0.02$. 
This matching definition is found to be fully efficient for leptons and close to 100\% 
efficient for jets. 

The response matrix is derived from MC simulations 
and has diagonal elements that are larger than 0.6 for all observables. 
The elements in the matrix are normalised
to the total number of reconstructed events determined from the sum of all
the bins associated with the x-axis. The response matrix is applied using two iterations,
which is found to provide convergence and avoid higher statistical uncertainties
in the tails of the corrected distributions.

Events that have no matched detector-level and particle-level pair are taken into account
by three factors that correct the detector-level observable distribution
to the particle-level observable distribution:
\begin{itemize}
\item The correction for events that pass the detector-level event selection 
but fail the particle-level event selection (\frnp);

\item The correction factor for events with a reconstructed 
pseudo-top-quark that has no counterpart at the particle level (\fmisassign);

\item The correction for events that fulfil the particle-level event selection requirements 
but fail the reconstruction-level event selection (\fpnr).

\end{itemize}

These correction factors are also derived from MC simulation.
The correction factors \frnp\ and \fmisassign\ 
are in the range between $0.65$--$0.70$ for all observables. 
The correction factor \fpnr{} is primarily dominated by the detector efficiency,
in particular by the $b$-tagging efficiency, and is in the
range $5.5$--$8.0$. They are found to be similar for the \ttbar\ generators discussed in section~\ref{sec:monte-carlo-samples}, 
such that no large MC modelling dependencies 
are observed between the different MC samples. 

The same unfolding procedure is applied to each of the distributions 
$\pTx(\pseudotop)$, $y(\pseudotop)$, $y(\pseudotopl\pseudotoph)$, $\pT(\pseudotopl\pseudotoph)$, 
and $m(\pseudotopl\pseudotoph)$.

The number of background events \Nbgnd\ and the correction factors \frnp\ and \fmisassign\ are 
functions of the detector-level pseudo-top-quark observables.  
The correction factor \fpnr\ is a function of the particle-level pseudo-top-quark observable \xparti. 
To evaluate the cross-section in bin $i$, it is also necessary to take account both the 
luminosity and bin width.

\subsection{Propagation of uncertainties}
\label{sec:propagation}

With the exception of the non-\ttbar\ backgrounds, each of the correction factors in eq.~(\ref{eqn:corrections}) is calculated using \powheg{}+\pythia{} events that are passed through the detector simulation.  The effect of the statistical uncertainty on \Mres\ is estimated by smearing the number of events in each element of the matrix, using a Poisson probability density function.  The statistical uncertainty on the correction factors (\frnp, \fmisassign\ and \fpnr) is evaluated by smearing the value in each bin using a Gaussian distribution following the statistical uncertainty in the bin. The correction factors and 
the response matrix are smeared simultaneously by performing 1000 pseudo-experiments
and repeating the unfolding procedure for each pseudo-experiment.
The statistical uncertainty for each measurement point is taken
from the root-mean-square  
($1 \sigma$)
of the spread of the unfolded distributions over the various
pseudo-experiments.

The statistical uncertainty on the reconstructed distributions in data 
is propagated to the final distributions by performing 1000 pseudo-experiments, following a Poisson distribution defined by the number of events in each bin $j$.  Similar to the MC statistical uncertainty, each bin of \xrecoj\ is independently fluctuated.

The experimental systematic uncertainty on the reconstructed distributions is evaluated by changing the values of the physics objects by their associated uncertainties.
The total uncertainty on the 
number of reconstructed
background events (\Nbgnd) is evaluated by summing in quadrature each of the background uncertainties discussed in section~\ref{sec:systematic_uncertainties}.

The systematic
uncertainty on the unfolded spectra due to the background is evaluated
by performing 1000 pseudo-experiments, following a normal
distribution with a width matching the total uncertainty band.  The
root-mean-square of the distribution of unfolded spectra
of the pseudo-experiments is taken as the uncertainty on the
background.  Due to the small number of background events and their generally small uncertainties, the background uncertainties are propagated in an uncorrelated manner. 
This choice reduces the impact of statistical fluctuations on the background uncertainty estimates 
for the final distributions.   

The systematic uncertainties on \Mres, \frnp, \fmisassign\ and \fpnr\ arising 
due to the choice of the \powheg{}+\pythia{} \ttbar\ MC model are each evaluated as a relative bias.
For a given MC simulation set
the bias is defined as the fully corrected unfolded yield for a 
given bin minus the true particle yield for that bin.  
The relative bias is defined as the difference in bias between the
nominal MC sample and the varied MC sample, both unfolded using the nominal MC sample.
For the \powheg{}+\pythia{} \ttbar\ MC sample, 
the bias is found to be consistent with zero within the statistical uncertainties of each 
measurement point.  For each \ttbar\ modelling systematic uncertainty, a pair of 
particle-level and detector-level spectra is generated.  One thousand pseudo-experiments are used to fluctuate
the reconstructed input spectrum within its statistical uncertainty. The pseudo-experiments are used to evaluate the statistical significance of the systematic variation in the output distribution. The relative bias is calculated for each pseudo-experiment.

The ISR/FSR systematic uncertainty 
is evaluated from the relative bias between
the \alpgen{}+\pythia{} central and shifted ISR/FSR samples.  The
uncertainty on the matrix-element calculation and matching scheme (the
generator uncertainty) is estimated from the relative bias of  \mcnlo{}+\herwig{} with respect to the \alpgen{}+\herwig{} \ttbar\ sample.  

Each of the \ttbar\ uncertainties is propagated individually and are then symmetrised before being combined, taking the larger of the upward or downward variation.

\subsection{Combination of lepton channels}
\label{sec:combination}

The electron and muon channel measurements of each pseudo-top-quark distribution are combined by using the Best Linear Unbiased Estimate
 (BLUE) method~\cite{Valassi2003391,Lyons:1988rp}.  The BLUE method determines the coefficients
(weights) to be used in a linear combination of the input measurements by minimising the total uncertainty of the combined
result.  All uncertainties are assumed to be distributed according to a Gaussian probability density function. The algorithm
takes both the statistical and systematic uncertainties and their correlations into account.  The BLUE combination was cross-checked against an average performed 
using the algorithm discussed in ref.~\cite{Aaron:2009bp}.  The result of the two methods are found to be consistent.
The MC statistical uncertainties on the correction factors for the two samples are assumed to be uncorrelated.  
The uncertainties related to the electron and muon efficiencies are also treated uncorrelated.
All other systematic uncertainties are treated as fully correlated. 
In particular, the total background systematic uncertainty is assumed to be completely correlated between the electron and muon
channel. The uncertainties
of the combined measurement are dependent on the observables but tend to closely follow the muon channel uncertainties.  

\section{Results}
\label{sec:results}

The measurements of the differential \ttbar\ cross-section corrected for detector effects
are presented as a function of the $\pT(\pseudotop)$ and $y(\pseudotop)$ for the hadronic or leptonic reconstruction of the pseudo-top-quarks, as well as the variables $y(\pseudotopl\pseudotoph)$, $\pT(\pseudotopl\pseudotoph)$ and $m(\pseudotopl\pseudotoph)$ of 
the pseudo-top-quark-pair system; \pseudotop{} refers to both the hadronic and leptonic pseudo-top-quark.  
The fiducial cross-section measurements are presented within the kinematic range defined in 
section~\ref{sec:meas-def}, and are evaluated as described in section \ref{sec:corr_chancomb}.

The measurements can be directly compared to Monte Carlo simulations that are generated
with top quarks with a mass of $172.5$~\GeV. The top quark mass dependence of the cross-sections
was derived with Monte Carlo samples generated with top quark masses from $165$ to $178$~\GeV.
The cross-sections depend linearily on the top mass.
The slope\footnote{The top mass ($m_{t}$) dependence of the cross-sections
can be obtained from the slope $\alpha$ quoted above via $\alpha \cdot (m_{t} - 172.5)$.}
is between $-0.1$ and $-0.3$ \%/\GeV{} for  $y(\pseudotop)$, $y(\pseudotopl\pseudotoph)$
and  $\pT(\pseudotopl\pseudotoph)$. For $m(\pseudotopl\pseudotoph)$ ($\pT(\pseudotop)$) it rises
from $-0.6$ ($-0.3$) \%/\GeV{} for low values to $0.4$ ($0.4$) \%/\GeV{} at high values.


The  \ttbar\ MC generators are of two types (for a detailed discussion see section~\ref{sec:monte-carlo-samples}): 
NLO matrix elements are used to describe the hard scattering $p p \to \ttbar$,
and LO multi-leg MC generators model \ttbar{} processes with up to five additional quark or gluon emissions.
All  \ttbar\ MC samples are normalised to the NNLO+NNLL QCD inclusive cross-section.   

Figures \ref{fig:lo_combination-t_pT}, \ref{fig:lo_combination-t_y} and \ref{fig:lo_combination-tt} show the corrected data for the differential 
variables noted above. Superimposed on the data are the expectations of the LO multi-leg MC generator \alpgen{} (see section~\ref{sec:monte-carlo-samples}):
\begin{itemize}
\item \alpgen{} interfaced with \pythia{} (\alpgen{}+\pythia{}),
\item \alpgen{} interfaced with \herwig{} (\alpgen{}+\herwig{}),
\item \alpgen{} interfaced with \pythia{} but with the renormalisation scale varied by a factor of 2 ($\alphas$ Up) and 
a factor of 0.5 ($\alphas$ Down). 
\end{itemize}

\begin{figure}[htbp] 
\centering
\subfigure[\label{fig:lo_combination-t_pt_had}]{\includegraphics[width=.49\linewidth]{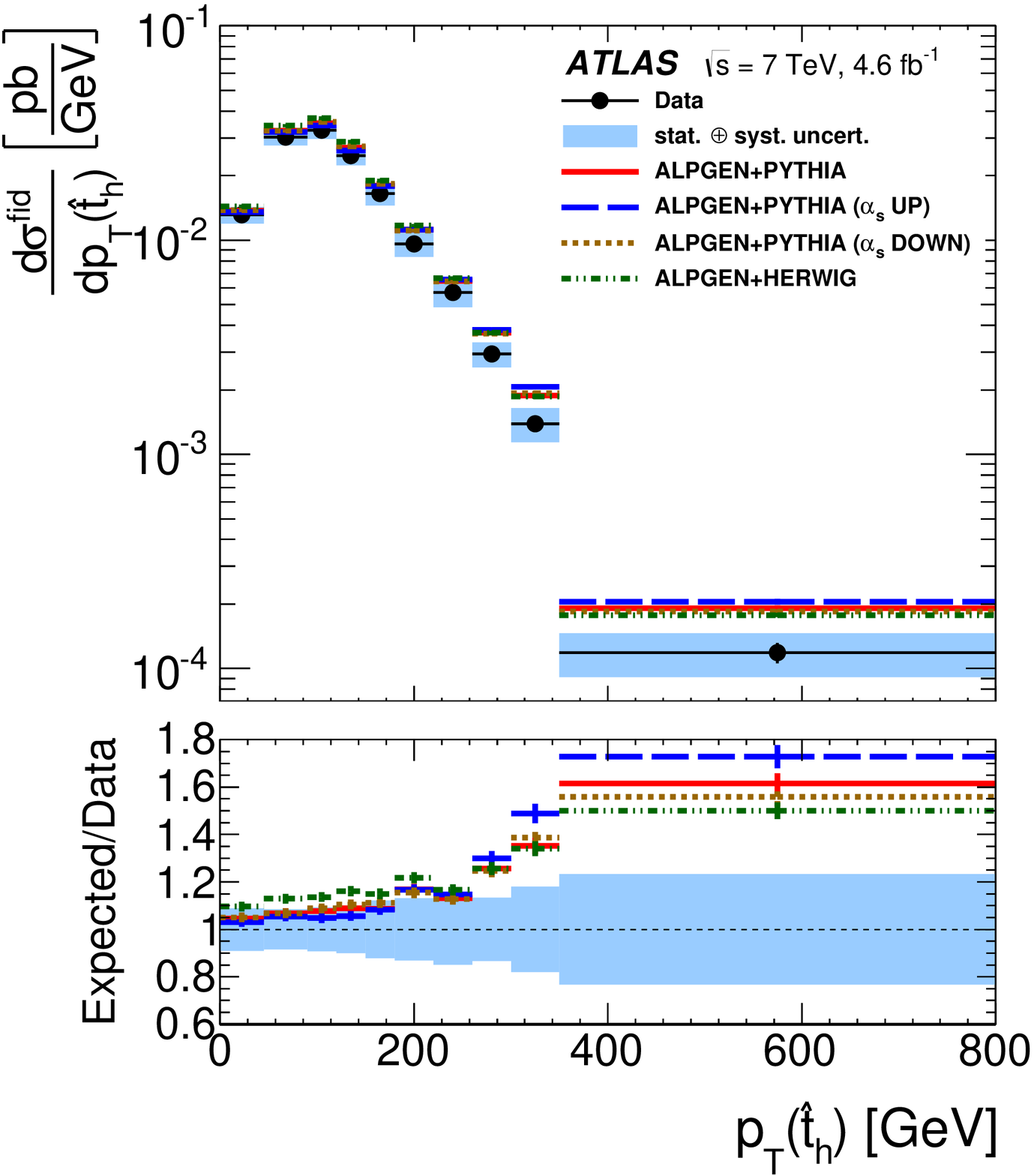}}
\subfigure[\label{fig:lo_combination-t_pt_lep}]{\includegraphics[width=.49\linewidth]{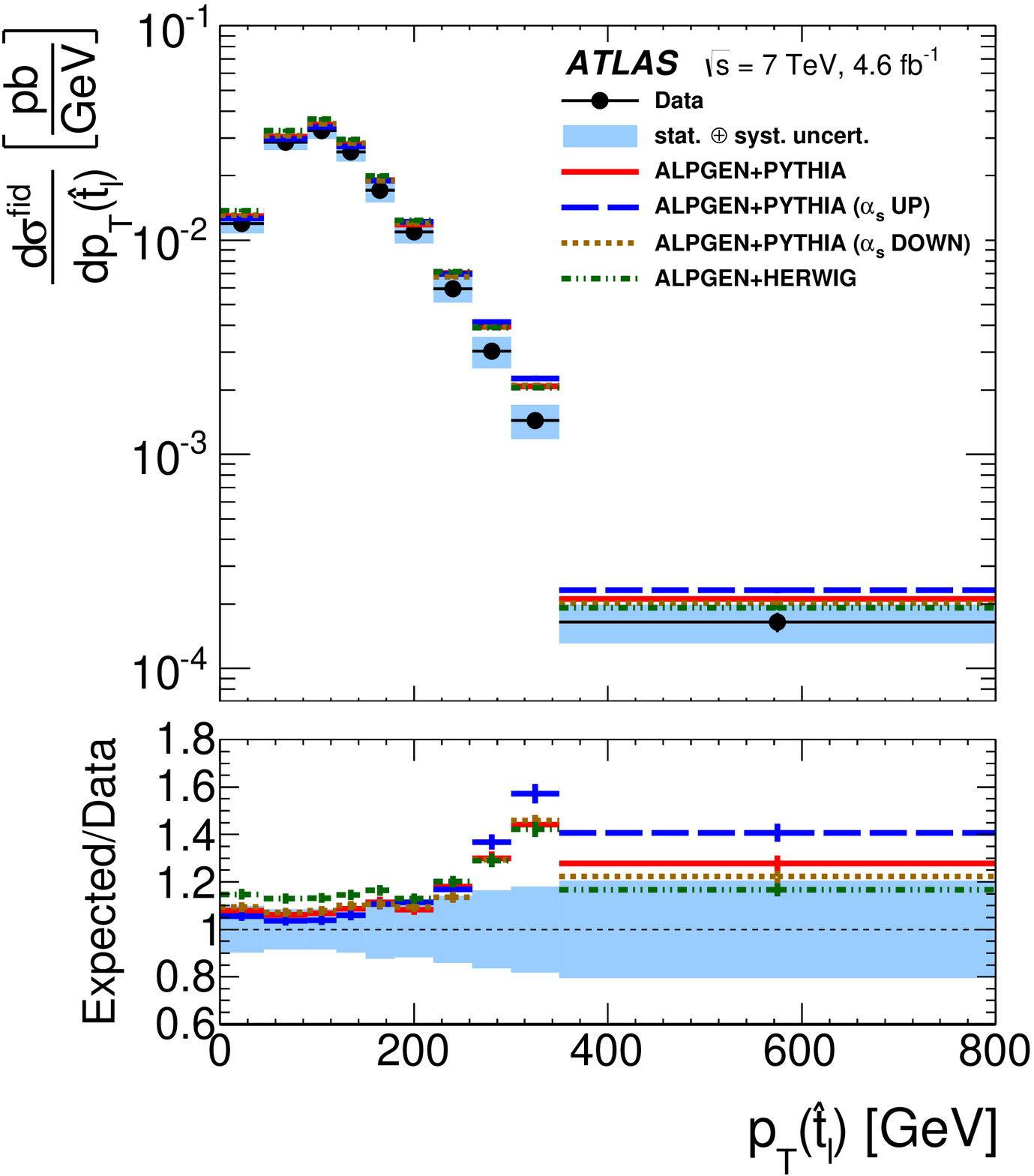}}
 \caption{Differential \ttbar\ cross-section after channel combination as a function of (a) the hadronic pseudo-top-quark $\pT(\pseudotoph)$ and (b) the leptonic pseudo-top-quark $\pT(\pseudotopl)$. The data points are shown with a blue band which represents the total uncertainty (statistical and systematic). The model predictions from several LO multi-leg MC generators described in the text are superimposed: the \alpgen{}+\herwig{} and the \alpgen{}+\pythia{} samples. For the \alpgen{}+\pythia{} sample, 
variations in the renormalisation scale by factors of 0.5 and 2.0 are also shown.}
\label{fig:lo_combination-t_pT}
\end{figure}

\begin{figure}[htbp] 
\centering
\subfigure[\label{fig:lo_combination-t_y_had}]{\includegraphics[width=.49\textwidth]{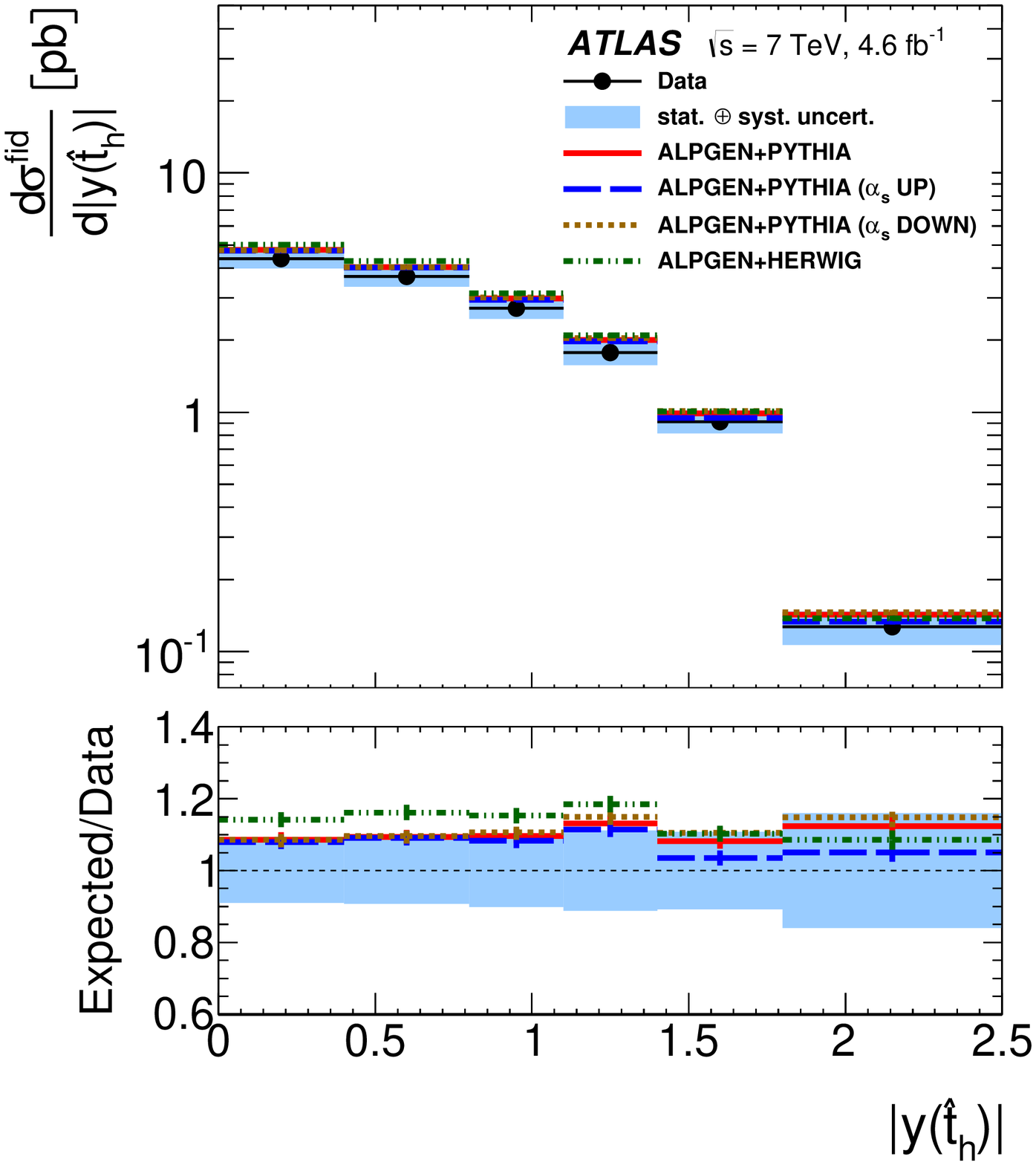}}
\subfigure[\label{fig:lo_combination-t_y_lep}]{\includegraphics[width=.49\textwidth]{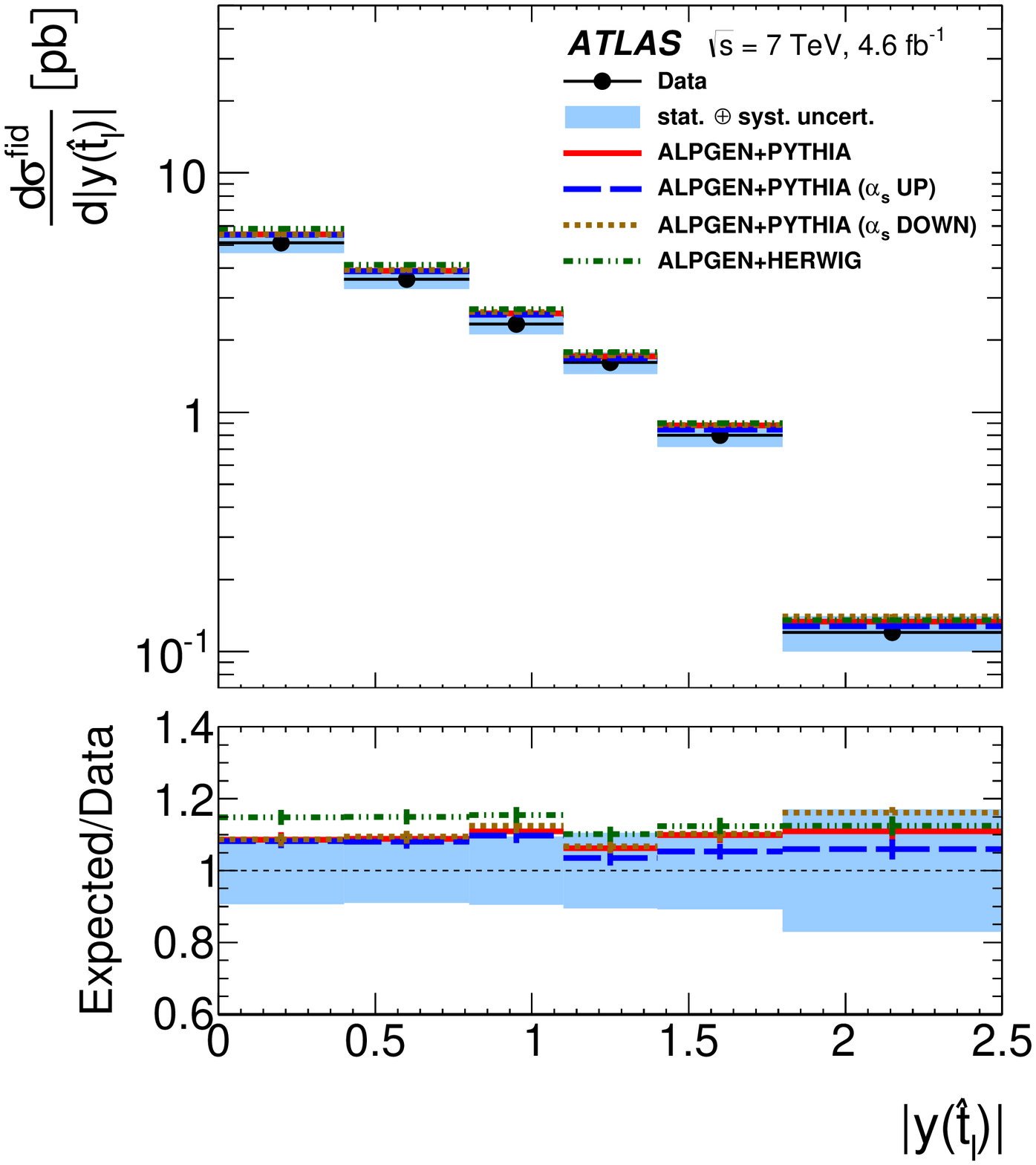}}
  \caption{Differential \ttbar\ cross-section after channel combination as a function of (a) the hadronic pseudo-top-quark rapidity $y(\pseudotoph)$ and (b) the leptonic pseudo-top-quark rapidity $y(\pseudotopl)$. The data points are shown with a blue band which represents the total uncertainty (statistical and systematic). The model predictions from several LO multi-leg MC generators described in the text are superimposed: the \alpgen{}+\herwig{} and the \alpgen{}+\pythia{} samples. For the \alpgen{}+\pythia{} sample, variations in the renormalisation scale by factors of 0.5 and 2.0 are also shown.}
\label{fig:lo_combination-t_y}
\end{figure}

\begin{figure}[htbp]
\centering
\subfigure[\label{fig:lo_combination-tt_pT}]{\includegraphics[width=.49\linewidth]{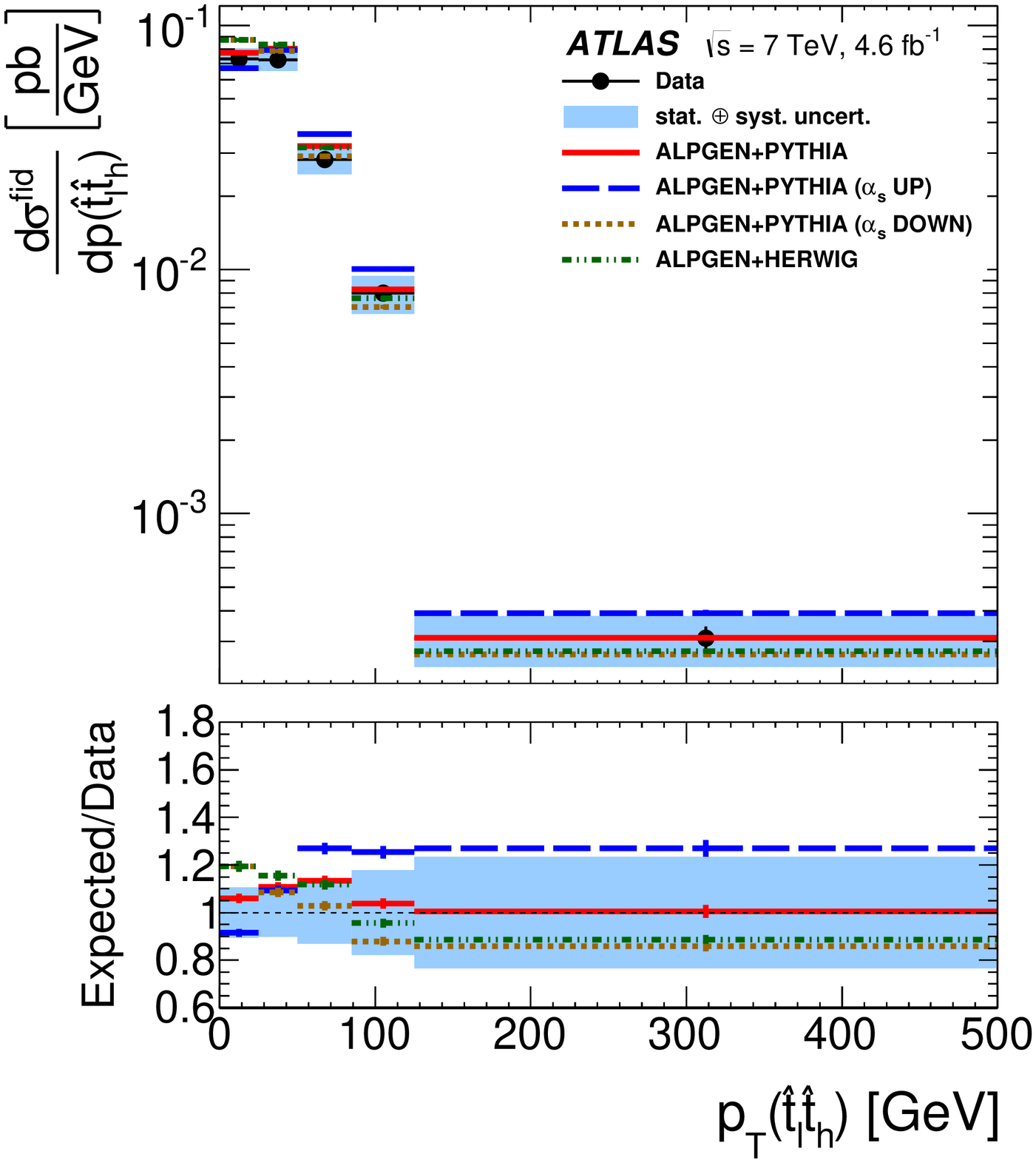}}
\subfigure[\label{fig:lo_combination-tt_y}]{\includegraphics[width=.49\linewidth]{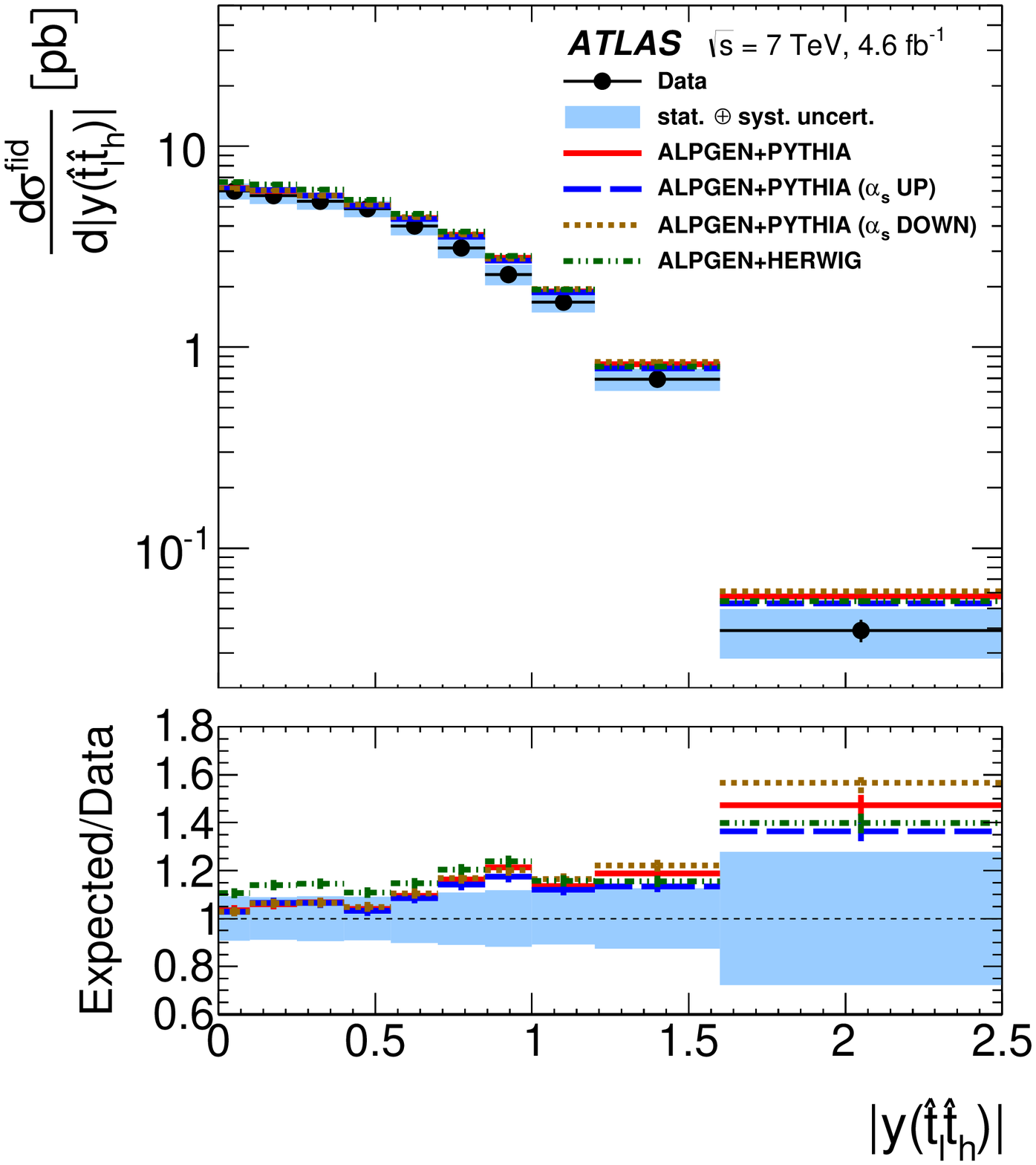}}
\subfigure[\label{fig:lo_combination-tt_m}]{\includegraphics[width=.49\linewidth]{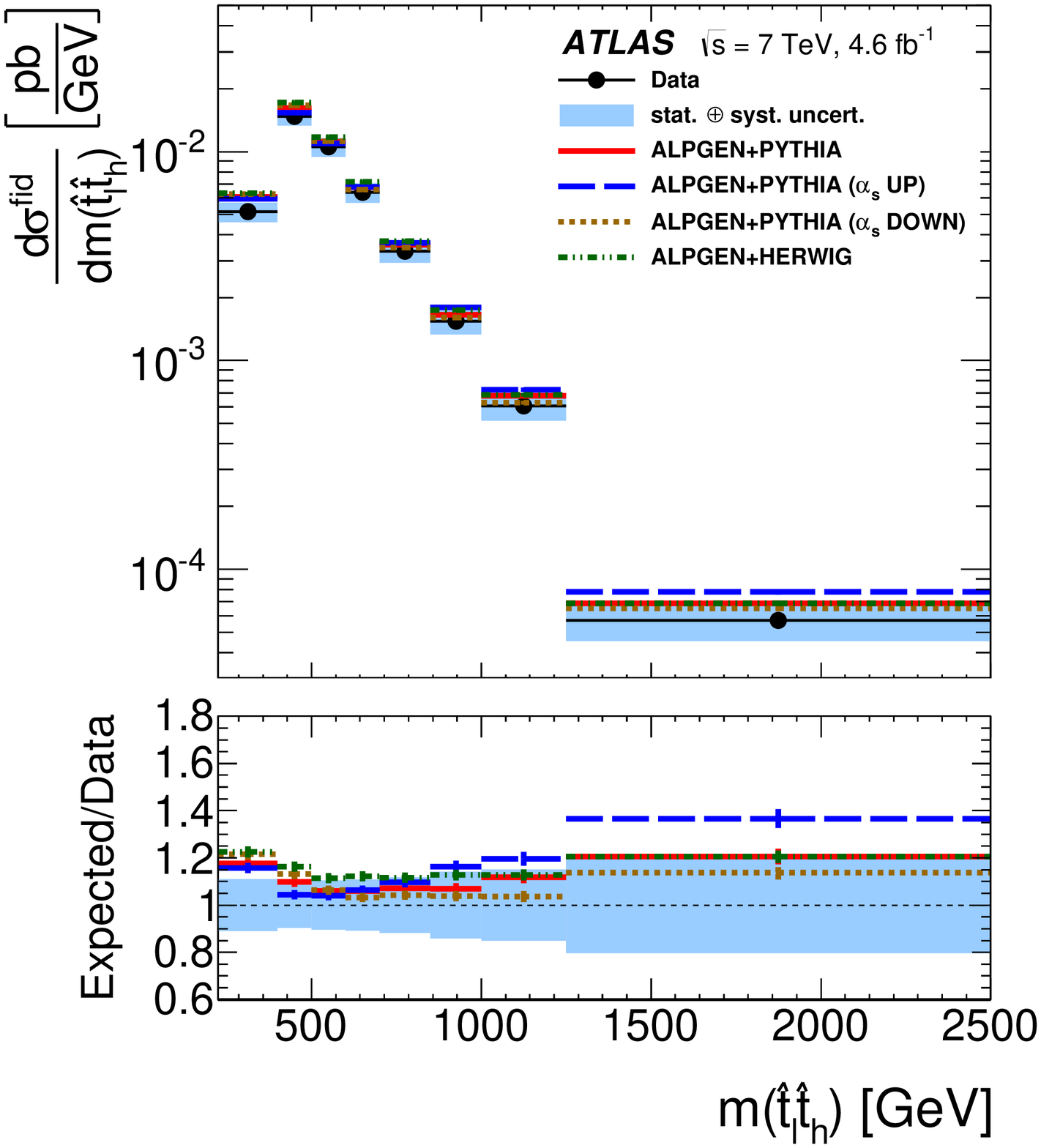}}
  \caption{Differential \ttbar\ cross-section after channel combination as a function of (a) the total leptonic and hadronic \ttbar\ pseudo-top-quark variable  $\pTx(\pseudotopl\pseudotoph)$, (b) the rapidity $y(\pseudotopl\pseudotoph)$ and (c) the mass $m(\pseudotopl\pseudotoph)$. The data points are shown with a blue band which represents the total uncertainty (statistical and systematic). The model predictions from several LO multi-leg MC generators described in the text are superimposed: the \alpgen{}+\herwig{} and the \alpgen{}+\pythia{} samples. For the \alpgen{}+\pythia{} sample, variations in the renormalisation scale by factors of 0.5 and 2.0 are also shown.}
\label{fig:lo_combination-tt}
\end{figure}

Both the \alpgen{}+\herwig{} and \alpgen{}+\pythia{} models indicate yields within the acceptance that are higher than the observed data. This is
evident in figures \ref{fig:lo_combination-t_y} and \ref{fig:lo_combination-tt_y} where for the complete rapidity range the models predict more events. 
The models also indicate a $\pTx(\pseudotop)$ spectrum that is harder than the ones observed in data, a possible consequence of
using the \cteq{} PDF set.  For the \alpgen{}+\pythia{} sample, the effect of 
increased or decreased radiation is illustrated with
the renormalisation scale changed by a factor of 2.0 (\alphas\ Up) and by a factor of 0.5 (\alphas\ Down) applied consistently to
 both \alpgen{} and \pythia{} as noted in section~\ref{sec:monte-carlo-samples}. 
The increased radiation gives fewer events at low $\pT(\pseudotop)$ and more at high $\pT(\pseudotop)$.
This is at the level of $5$--$10$\%.  For the leptonic $\pT(\pseudotop)$ this effect is slightly larger as shown in figure \ref{fig:lo_combination-t_pt_lep}.
The $\pT(\pseudotopl\pseudotoph)$ distribution, 
shown in figure~\ref{fig:lo_combination-tt_pT}, is very sensitive to additional radiation. 
When the renormalisation scale factor changes from $0.5$ to $2.0$,
$15$--$20$\% more events are observed at low $\pT(\pseudotopl\pseudotoph)$ 
and $20$--$30$\% fewer  at high $\pT(\pseudotopl\pseudotoph)$. Increased radiation also
leads to $5$\% fewer events at low $m(\pseudotopl\pseudotoph)$ and $10$\% more events at high $m(\pseudotopl\pseudotoph)$ as shown in figure \ref{fig:lo_combination-tt_m}.
Nevertheless, the (\alphas\ Down) variation is not sufficient to restore agreement between the data and the MC simulation.
The \alpgen+\herwig{} sample follows the \alpgen+\pythia{} sample with the (\alphas\ Down) variation.

Figures \ref{fig:nlo_combination-t_pT}, \ref{fig:nlo_combination-t_y} and \ref{fig:nlo_combination-tt} compare the data with expectations of the NLO MC generators \powheg\ and \mcnlo. 
In particular, the following NLO variants are shown:
\begin{itemize} 
\item \powheg{}-hvq v.4 generator with the \ctten{} PDF set interfaced with \pythia{} using the ``C" variant of the Perugia 2011 tunes and
\cteq{} PDF set (\powheg{}+\pythia{}).
\item \powheg{}-hvq v.4 generator where the \ctten{} PDF set is replaced with the\newline \hera{} PDF set to assess the sensitivity of the distributions to 
changes in the gluon PDF (\powhegwith{herapdf}+\pythia{}). 
\item \mcnlo{} generator with the  \ctten{} PDF set interfaced with \herwig{} and \jimmy{} with the AUET2 tune (\mcnlo{}+\herwig{}). 
\end{itemize}

\begin{figure}[htbp] 
\centering
\subfigure[\label{fig:nlo_combination-t_pt_had}]{\includegraphics[width=.49\linewidth]{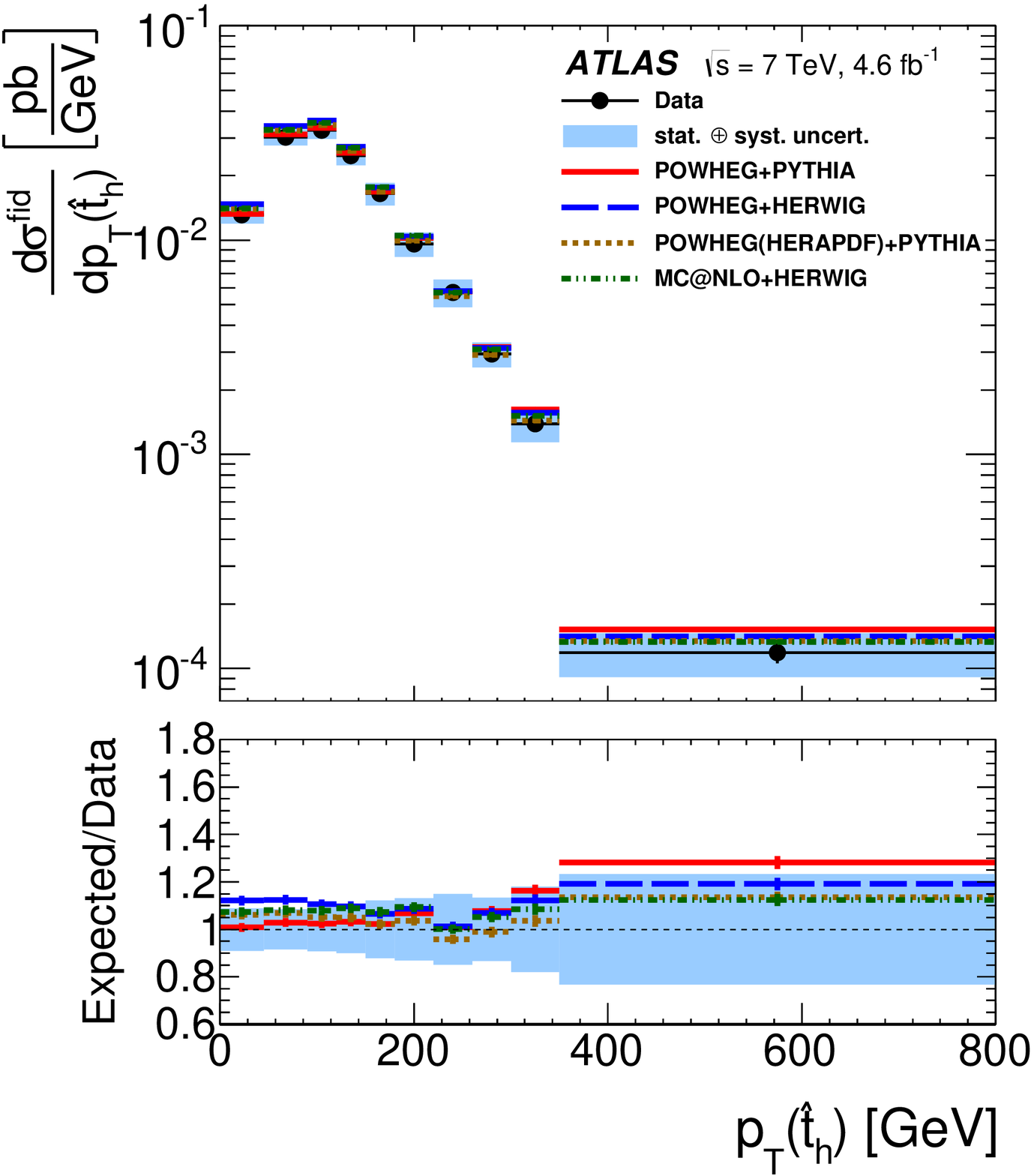}}
\subfigure[\label{fig:nlo_combination-t_pt_lep}]{\includegraphics[width=.49\linewidth]{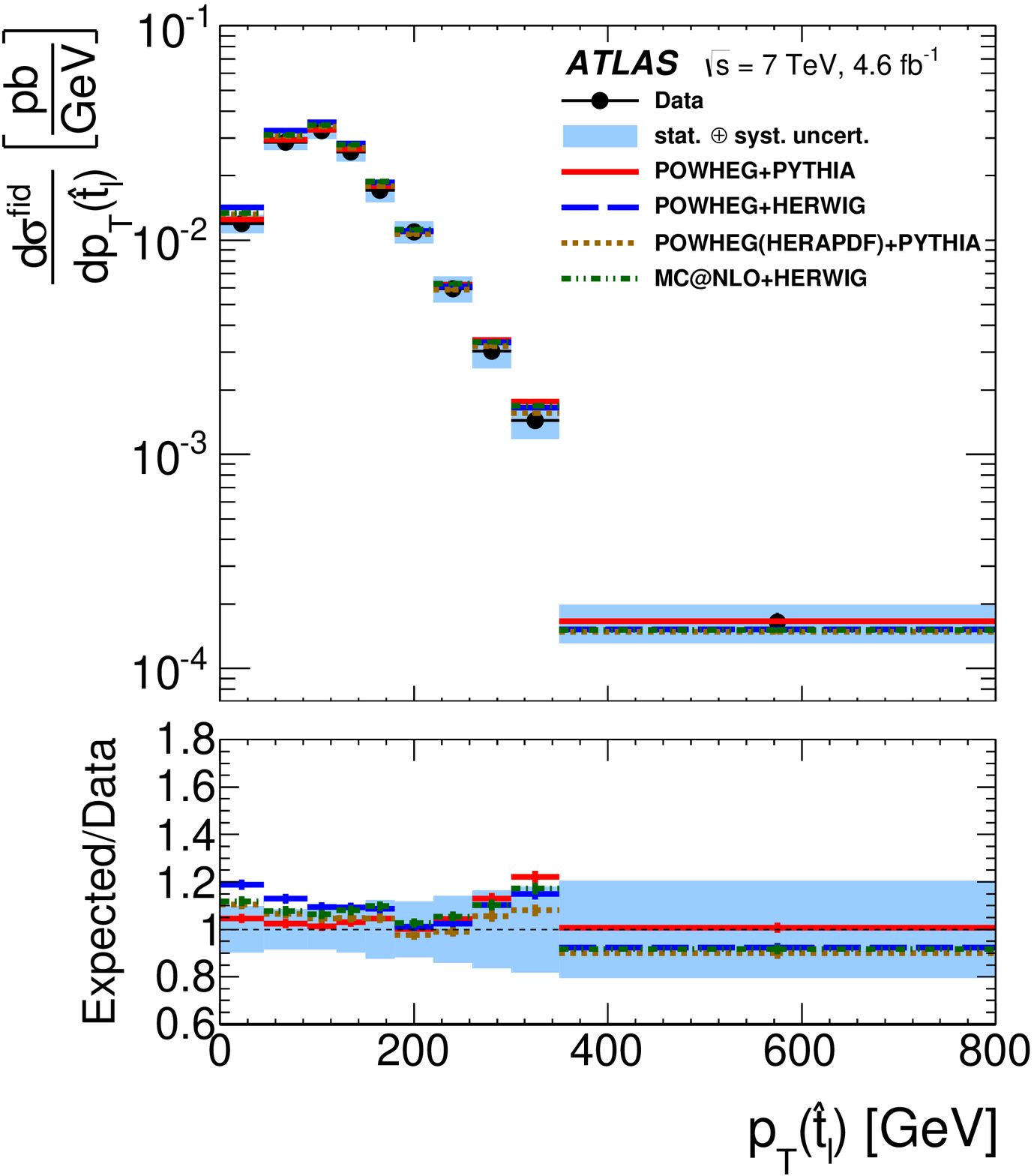}}
 \caption{Differential \ttbar\ cross-section after channel combination as a function of (a) the hadronic 
 pseudo-top-quark~$\pTx(\pseudotoph)$ and (b) the leptonic pseudo-top-quark~$\pTx(\pseudotopl)$. The data points are shown with a blue band which represents the total uncertainty (statistical and systematic). The model predictions from several NLO MC generators described in the text are superimposed: \powhegwith{ct10}+\pythia, 
\powhegwith{herapdf}+\pythia, \powheg+\herwig\ and \mcnlo+\herwig.}
\label{fig:nlo_combination-t_pT}
\end{figure}

\begin{figure}[htbp] 
\centering
\subfigure[\label{fig:nlo_combination-t_y_had}]{\includegraphics[width=.49\textwidth]{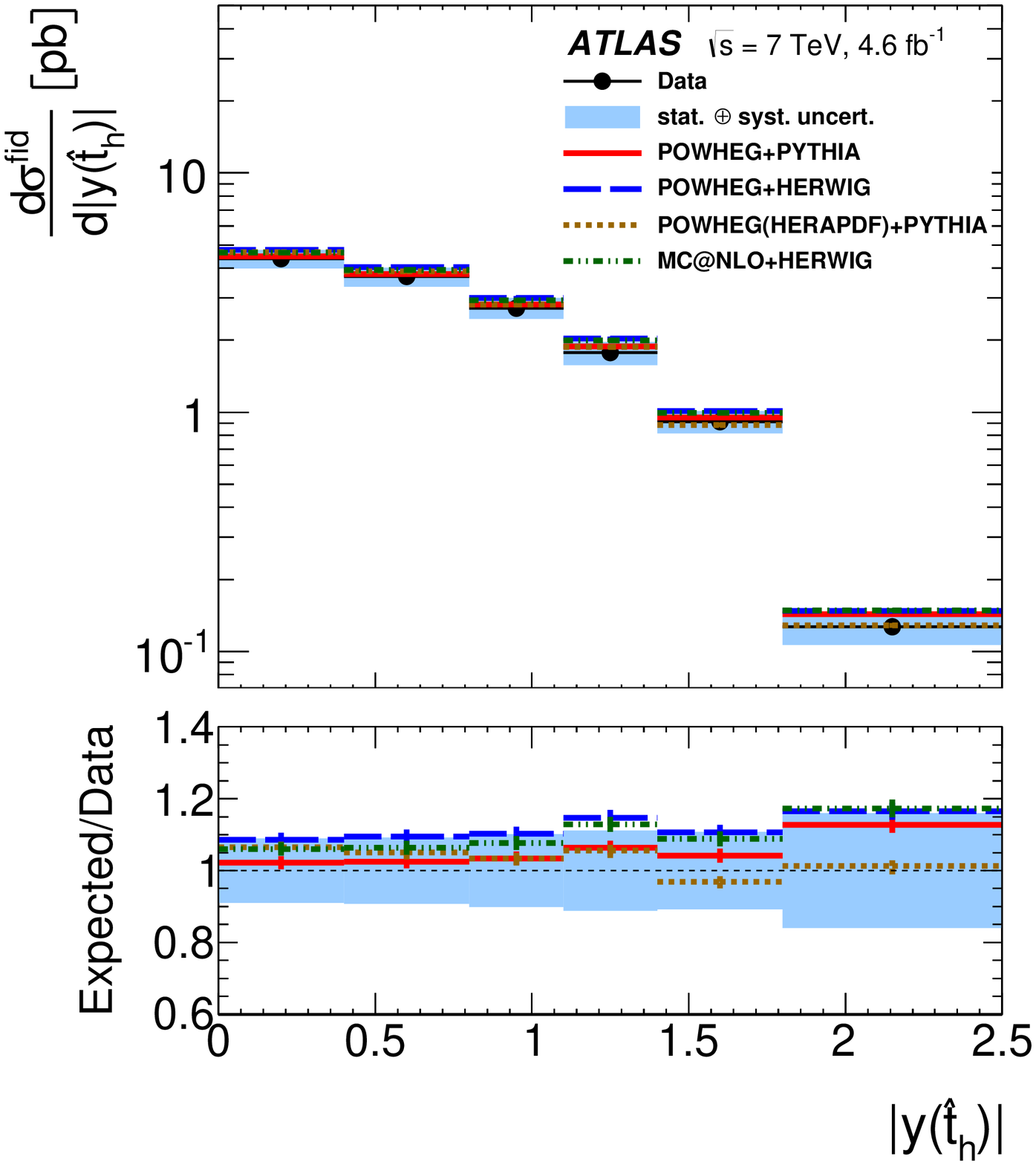}}
\subfigure[\label{fig:nlo_combination-t_y_lep}]{\includegraphics[width=.49\textwidth]{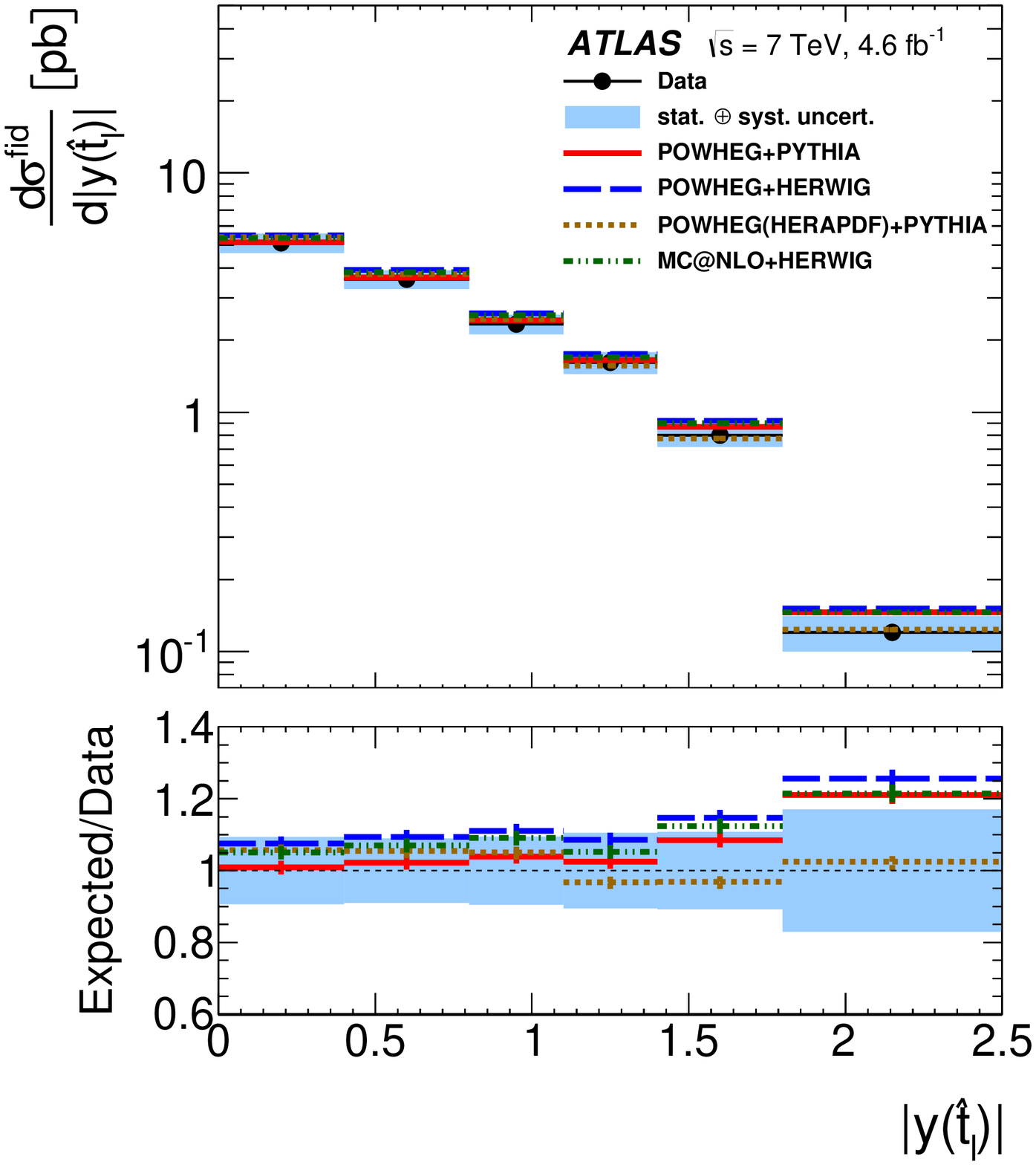}}
  \caption{Differential \ttbar\ cross-section after channel combination as a function of (a) the hadronic pseudo-top-quark rapidity~$y(\pseudotoph)$ and (b) the leptonic pseudo-top-quark rapidity~$y(\pseudotopl)$. The data points are shown with a blue band which represents the total uncertainty (statistical and systematic). The model predictions from several NLO MC generators described in the text are superimposed: \powhegwith{ct10}+\pythia, \powhegwith{herapdf}+\pythia, \powheg+\herwig\ and \mcnlo+\herwig.}
\label{fig:nlo_combination-t_y}
\end{figure}

\begin{figure}[htbp]
\centering
\subfigure[\label{fig:nlo_combination-tt_pT}]{\includegraphics[width=.49\linewidth]{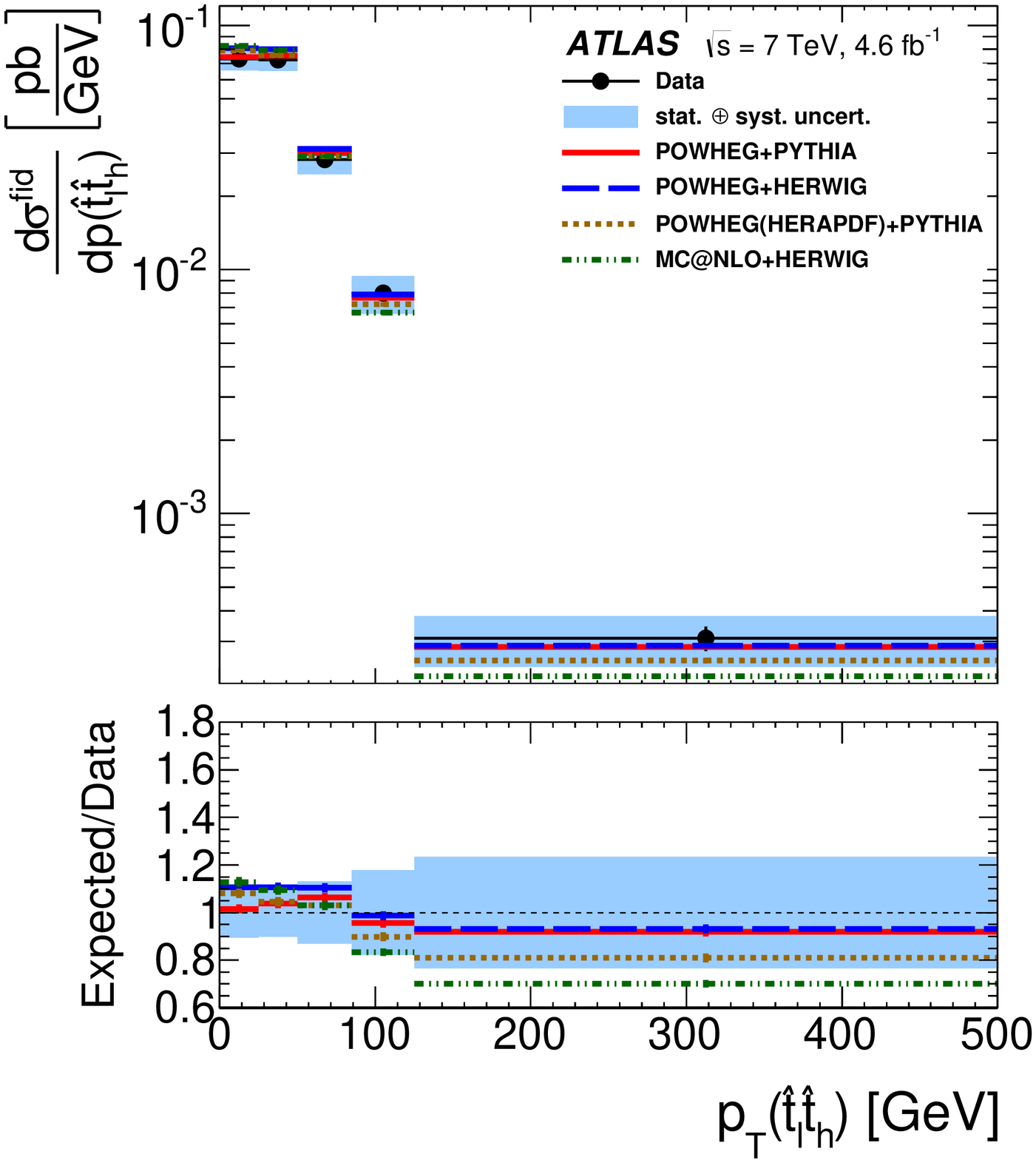}}
\subfigure[\label{fig:nlo_combination-tt_y}]{\includegraphics[width=.49\linewidth]{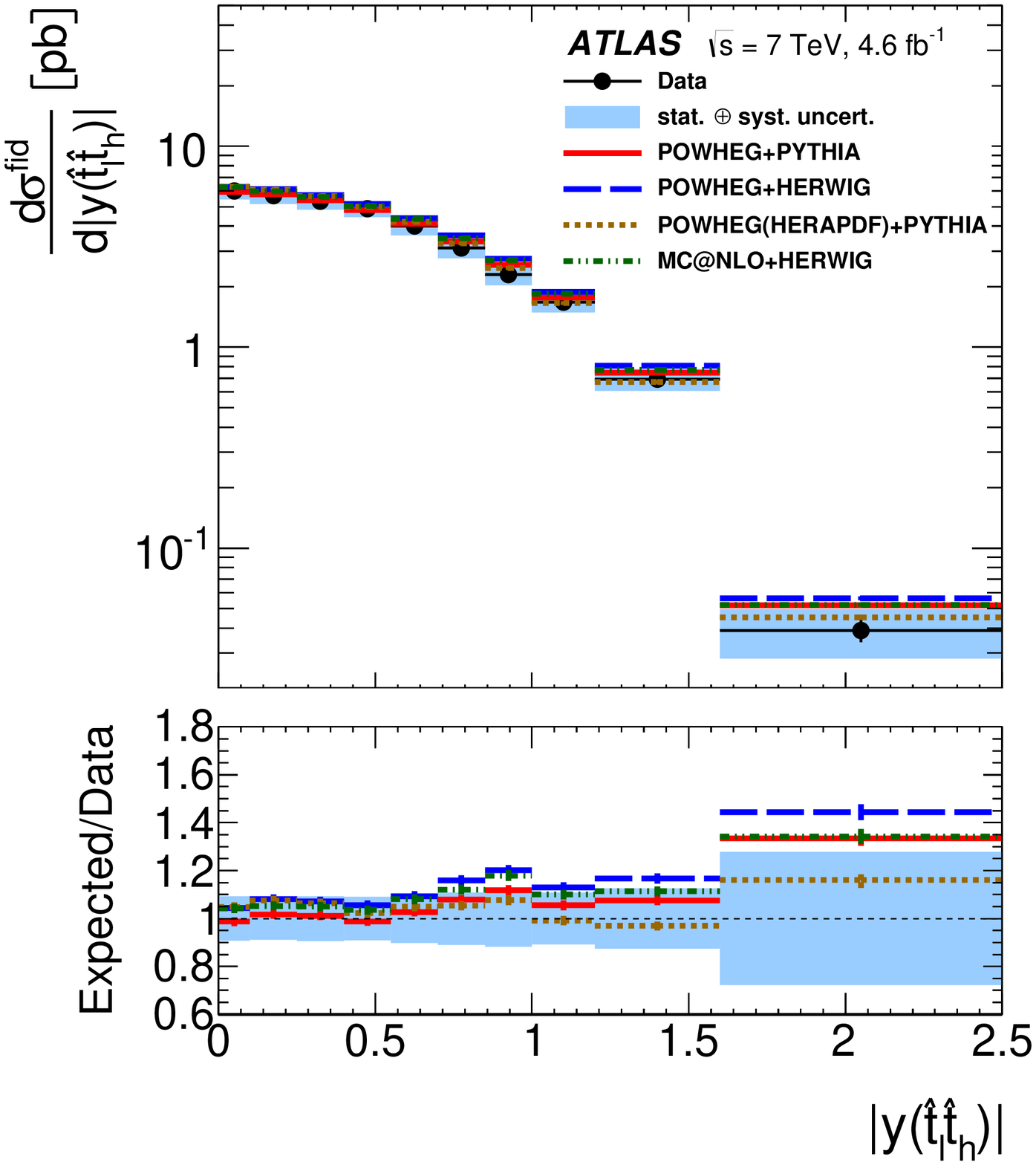}}
\subfigure[\label{fig:nlo_combination-tt_m}]{\includegraphics[width=.49\linewidth]{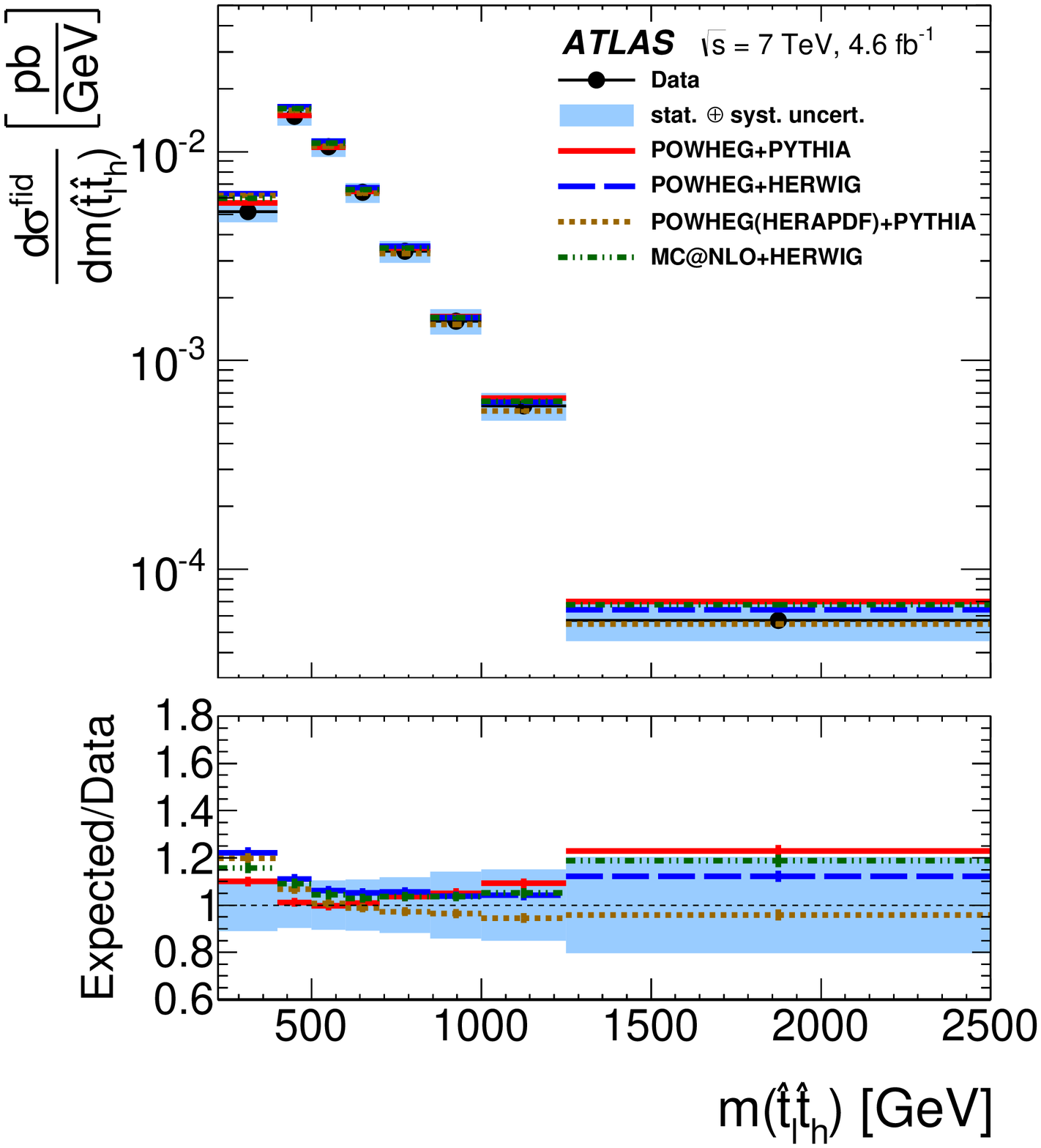}}
  \caption{Differential \ttbar\ cross-section after channel combination as a function of (a) the total leptonic and hadronic \ttbar\ pseudo-top-quark variables  $\pTx(\pseudotopl\pseudotoph)$, (b) the rapidity $y(\pseudotopl\pseudotoph)$ and (c) the mass $m(\pseudotopl\pseudotoph)$. The data points are shown with a blue band which represents the total uncertainty (statistical and systematic). The model predictions from several NLO MC generators described in the text are superimposed: \powhegwith{ct10}+\pythia, \powhegwith{herapdf}+\pythia, \powheg+\herwig\ and \mcnlo+\herwig.}
\label{fig:nlo_combination-tt}
\end{figure}

The individual hadronic and leptonic $y(\pseudotop)$ distributions
in figure~\ref{fig:nlo_combination-t_y} are well described by each of the 
NLO MC models. The \powhegwith{herapdf}+\pythia\ sample is
the one closest to predicting the data, whereas the other NLO models predict a more forward distribution.  This is also
true for the hadronic and leptonic $\pT(\pseudotop)$ 
distributions (see figure \ref{fig:nlo_combination-t_pT}) 
for which the \powhegwith{herapdf}+\pythia{} sample lowers the cross section at high $\pT(\pseudotop)$ 
with respect to the nominal sample, which results in a better description of the data.
For these variables \mcnlo\ also gives a good description.
The $\pT(\pseudotopl\pseudotoph)$ distribution, shown in figure \ref{fig:nlo_combination-tt_pT}, highlights the different hard-gluon emission models. 
The \mcnlo\ prediction is lower than the data at high $\pT(\pseudotopl\pseudotoph)$, 
as expected due to the softer fifth-jet \pT\ from \mcnlo{} in comparison to the other generators~\cite{Aad:2014iaa}. 

All the models predict an excess in the lower $m(\pseudotopl\pseudotoph)$ region, as shown in figure \ref{fig:nlo_combination-tt_m}, 
implying that the threshold region description is inadequate. The 
\powhegwith{herapdf}+\pythia{} sample
agrees well with the high-mass $m(\pseudotopl\pseudotoph)$ tail, while the other samples overestimate the tail. 
This is consistent with the softer gluon component in the \hera{} PDF set compared to the one in the  \ctten{} PDF set. 
The $y(\pseudotopl\pseudotoph)$ distribution is reasonably predicted 
for all models in the low $y(\pseudotopl\pseudotoph)$ region, but only the \powheg+\pythia\ model with the \hera{} PDF set provides a good overall description. 
The $\pTx(\pseudotopl\pseudotoph)$ spectrum is sensitive to the
extra radiation produced in the parton collision process. 
All models agree with the data within the systematic uncertainties.
If these uncertainties can be reduced, this suggests 
that the $\pTx(\pseudotopl\pseudotoph)$ distribution can be used to constrain phenomenological radiation parameters in future MC tunes.

\FloatBarrier
\section{Conclusions}
\label{sec:conclusions}

Differential fiducial \ttbar\ cross-section measurements are presented for kinematic variables of the pseudo-top-quark
(\pseudotop), defined at the particle level by the decay products of the $W$ boson
and $b$-quark occurring in top-quark decays. 

The pseudo-top-quark approach is a new tool to probe QCD in the top-quark sector.
It is an experimental observable that is 
correlated with the top-quark parton and 
is used to define differential \ttbar{} cross-sections with reduced model dependence. 
It can also be used to assess how well MC simulations can describe the \ttbar{} production 
mechanism in proton--proton collisions and to compare various MC models.

The present measurements were performed within a kinematic range that closely matches 
the acceptance of the reconstructed objects and each reconstructed kinematic variable 
distribution is corrected (unfolded) for the effects of detector efficiency and resolution. 

The differential fiducial cross-sections are measured with the ATLAS detector at the LHC
for proton--proton collisions at $\sqrt{s}=7$~\TeV\ and an integrated luminosity of 4.6~\ifb, as a function of $\pTx(\pseudotoph)$, $y(\pseudotoph)$, $\pTx(\pseudotopl)$, $y(\pseudotopl)$, 
$y(\pseudotopl\pseudotoph)$, 
$\pTx(\pseudotopl\pseudotoph)$, and $m(\pseudotopl\pseudotoph)$, where \pseudotopl{} and \pseudotoph{}
refers to the hadronic or leptonic pseudo-top-quark.  The distributions provide complementary information and show some sensitivity to the selected PDF set, the parton shower procedure, and 
to a lesser extent the matrix element and parton shower
matching scheme.  The larger acceptance for the \alpgen\ MC models resulted in an excess of events reconstructed within the
fiducial region. The higher \alphas\ variation is seen to be disfavoured at higher 
$\pT(\pseudotopl\pseudotoph)$ and $m(\pseudotopl\pseudotoph)$ values.  Among the several \ttbar~MC models used, 
the  \powhegwith{herapdf}+\pythia{} sample provides the best representation for all of the distributions, 
except for low $m(\pseudotopl\pseudotoph)$ where many of the MC models predict a higher cross-section than the data at low mass values.  The \mcnlo\ prediction is seen to produce a recoil distribution that is too soft with respect to the data.  
These measurements are currently limited by the systematic uncertainty, the main components being the $b$-tagging uncertainty, the 
jet energy measurement uncertainty and the modelling uncertainty of the initial and final state parton showers.
\section*{Acknowledgements}


We thank CERN for the very successful operation of the LHC, as well as the
support staff from our institutions without whom ATLAS could not be
operated efficiently.

We acknowledge the support of Anapests, Argentina; YerPhI, Armenia; ARC,
Australia; BMWFW and FWF, Austria; ANAS, Azerbaijan; SSTC, Belarus; CNPq and FAPESP,
Brazil; NSERC, NRC and CFI, Canada; CERN; CONICYT, Chile; CAS, MOST and NSFC,
China; COLCIENCIAS, Colombia; MSMT CR, MPO CR and VSC CR, Czech Republic;
DNRF, DNSRC and Lundbeck Foundation, Denmark; EPLANET, ERC and NSRF, European Union;
IN2P3-CNRS, CEA-DSM/IRFU, France; GNSF, Georgia; BMBF, DFG, HGF, MPG and AvH
Foundation, Germany; GSRT and NSRF, Greece; ISF, MINERVA, GIF, I-CORE and Benoziyo Center,
Israel; INFN, Italy; MEXT and JSPS, Japan; CNRST, Morocco; FOM and NWO,
Netherlands; BRF and RCN, Norway; MNiSW and NCN, Poland; GRICES and FCT, Portugal; MNE/IFA, Romania; MES of Russia and ROSATOM, Russian Federation; JINR; MSTD,
Serbia; MSSR, Slovakia; ARRS and MIZ\v{S}, Slovenia; DST/NRF, South Africa;
MINECO, Spain; SRC and Wallenberg Foundation, Sweden; SER, SNSF and Cantons of
Bern and Geneva, Switzerland; NSC, Taiwan; TAEK, Turkey; STFC, the Royal
Society and Leverhulme Trust, United Kingdom; DOE and NSF, United States of
America.

The crucial computing support from all WLCG partners is acknowledged
gratefully, in particular from CERN and the ATLAS Tier-1 facilities at
TRIUMF (Canada), NDGF (Denmark, Norway, Sweden), CC-IN2P3 (France),
KIT/GridKA (Germany), INFN-CNAF (Italy), NL-T1 (Netherlands), PIC (Spain),
ASGC (Taiwan), RAL (UK) and BNL (USA) and in the Tier-2 facilities
worldwide.

\newpage
\bibliographystyle{JHEP}
\bibliography{TOPQ_2013_07}

\clearpage
\input{atlas_authlist}
\end{document}

%% file: atlas_authlist.tex
\begin{flushleft}
{\Large The ATLAS Collaboration}

\bigskip

G.~Aad$^{\rm 85}$,
B.~Abbott$^{\rm 113}$,
J.~Abdallah$^{\rm 152}$,
S.~Abdel~Khalek$^{\rm 117}$,
O.~Abdinov$^{\rm 11}$,
R.~Aben$^{\rm 107}$,
B.~Abi$^{\rm 114}$,
M.~Abolins$^{\rm 90}$,
O.S.~AbouZeid$^{\rm 159}$,
H.~Abramowicz$^{\rm 154}$
% ...
\bigskip
\\
% ...
$^{11}$ Institute of Physics, Azerbaijan Academy of Sciences, Baku, Azerbaijan\\
% ...
$^{85}$ CPPM, Aix-Marseille Universit{\'e} and CNRS/IN2P3, Marseille, France\\
% ...
$^{90}$ Department of Physics and Astronomy, Michigan State University, East Lansing MI, United States of America\\
% ...
$^{107}$ Nikhef National Institute for Subatomic Physics and University of Amsterdam, Amsterdam, Netherlands\\
% ...
$^{113}$ Homer L. Dodge Department of Physics and Astronomy, University of Oklahoma, Norman OK, United States of America\\
$^{114}$ Department of Physics, Oklahoma State University, Stillwater OK, United States of America\\
% ...
$^{117}$ LAL, Universit{\'e} Paris-Sud and CNRS/IN2P3, Orsay, France\\
% ...
$^{152}$ Institute of Physics, Academia Sinica, Taipei, Taiwan\\
% ...
$^{154}$ Raymond and Beverly Sackler School of Physics and Astronomy, Tel Aviv University, Tel Aviv, Israel\\
% ...
$^{159}$ Department of Physics, University of Toronto, Toronto ON, Canada\\
% ...
\end{flushleft}
